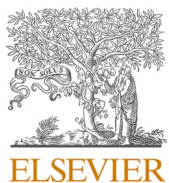
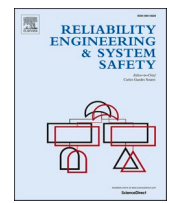

# Algorithmic analysis of a complex reliability system subject to multiple events with a preventive maintenance strategy and a Bernoulli vacation policy through MMAPs

Juan Eloy Ruiz-Castro [a], Hugo Alaín Zapata-Ceballos [b]

[a] *Dept. of Statistics and Operations Research and IMAG, University of Granada, Spain*
[b] *Dept. of Mathematics, University of Sucre, Colombia*

A B S T R A C T

In this work, a single-unit multi-state system is considered. The system is subject to internal failures, as well as external shocks with multiple consequences. It also incorporates a preventive maintenance strategy and a Bernoulli vacation policy for the repairperson. It is algorithmically modeled in both continuous and discrete time using Marked Markovian Arrival Processes (MMAP). The system's operation/degradation level is divided into an indeterminate number of levels. Upon returning from a vacation period, the repair technician may initiate corrective repair, perform preventive maintenance, replace the unit, remain idle at the workplace, or begin a new vacation period. The decision in the latter two cases is made probabilistically based on the system's operational level. This methodology allows the model and its associated measures to be algorithmically derived in both transient and stationary regimes, presented in a matrix-algorithmic form. Analytical-matrix methods are used to obtain the system's steady-state behaviour as well as various performance measures. Costs and rewards are introduced to analyze when the system becomes profitable. Measures associated with costs over time and in the stationary regime are defined and considered for optimization studies. A numerical example demonstrates the versatility of the model by solving a probabilistic optimization problem using a multi-objective Pareto analysis approach and performing a comparative evaluation of multiple models. Genetic algorithm is applied to find the optimization results in the reduced solution space. All modeling and associated measures have been computationally implemented in Matlab.

## 1. Introduction

Multistate systems play a crucial role in modern reliability analysis as they provide a more realistic representation of the behaviour of complex systems. Traditional reliability theory considers that both systems and their components can exist in only two operational states: either fully functional or completely failed. However, this binary approach has evolved with the introduction of multi-state systems (MSS). A multi-state system (MSS) is a complex and advanced system capable of functioning at various performance levels and encountering multiple failure modes, each exerting a unique influence on the system's overall operation. These systems possess a finite number of performance states, with each failure mode impacting their behaviour differently. As complex engineered systems continue to advance, developments in sensing technologies have uncovered that systems and their components often pass through multiple distinct states of degradation over time. As a result, the integration of multi-state systems into reliability analysis, moving beyond the traditional binary framework, has become an established and essential practice.

The literature in the field of multistate models is extensive, reflecting a wide range of reliability studies that address various aspects such as system structure, performance levels, degradation processes, and maintenance strategies. The earliest research on multi-state system (MSS) reliability dates back to Hirsch et al. [1]. Murchland [2] further explored the concept, which has undergone extensive development since then. From 1975 onward, significant advancements have been made in the study and modeling of MSS reliability [3]. Key foundational concepts of MSS were established by Murchland [2], El-Neweihi et al. [4], Barlow and Wu [5], and Ross [6], including the definition of the structure function for coherent multi-state systems.

In recent years, there has been a considerable increase in studies modeling multi-state reliability systems with optimization problems.

Thus, Gu et al. [7] addressed the reliability optimization of a multi-state system featuring two performance-sharing groups. The primary goal is to efficiently determine the optimal configuration policy that maximizes system reliability. The approach leverages stochastic ordering techniques and employs a genetic algorithm to obtain optimal solutions within a reduced solution space.

Another fundamental aspect of reliability is the implementation of effective maintenance policies. These policies play a crucial role in preventing significant economic losses and personal harm by ensuring that equipment and systems operate safely and efficiently over time.

There are many types of maintenance policies, each designed to address different operational needs and risk levels. Some of the most common include preventive maintenance, which involves regular inspections and servicing to avoid failures; predictive maintenance, which uses data and monitoring tools to anticipate issues before they occur; and corrective maintenance, which is carried out after a fault has been detected in order to restore functionality. Each of these strategies contributes to improving system reliability and minimizing unexpected downtime. In addition to maintenance policies, there are also methodologies designed to enhance system reliability, such as the use of redundant systems. These systems involve duplicating critical components or functions so that, in the event of a failure, operations can continue without interruption. Some relevant references have emerged in recent years. For instance, the study by Zhao et al. [8] compares replacement policies based on periodic intervals and a predetermined number of repairs from a cost rate perspective. These two maintenance strategies are analytically evaluated through the optimization of integrated models. A cold standby system is modeled using semi-Markov processes, where component lifetimes do not follow exponential distributions and partial repairability is taken into account, as presented by Li et al. [9]. An approximation algorithm is proposed and validated to solve the integral equations associated with the model. Recently, Levitin et al. [10] developed a corrective maintenance policy to restore the performance of the production subsystem while minimizing maintenance costs. To this end, they propose a new numerical algorithm that determines the optimal corrective maintenance policy by solving the optimization problem using a genetic algorithm. In the same year, Levitin et al. [11] proposed a numerical algorithm to assess mission success probability under a maintenance policy and employed a genetic algorithm to identify the optimal corrective maintenance policy. Wu et al. [12] derived, through detailed mathematical formulations, the expected cost functions necessary for determining optimal replacement policies in discrete-time replacement models. It considers random opportunities for preventive replacements in addition to scheduled or failure-driven replacements. The study analyzes two replacement disciplines and evaluates six distinct models based on replacement priority criteria.

Another interesting aspect of reliability is the incorporation of vacation policies into the system in a way that minimizes costs. Properly aligning workforce availability with maintenance and operational demands can significantly enhance system efficiency and reduce unnecessary expenses. A random vacation period is defined as the time during which the repairperson is unavailable in a repair channel. The repairperson may become deliberately unavailable for a certain period due to various reasons, such as minimizing idle periods or utilizing their efficiency for secondary services. Various strategies can be introduced for repairperson vacation periods, including the *N*-policy, multiple and single vacation policies, Bernoulli vacation policy, multiple and single working vacation policies, and vacation interruption policies, among others.

It is crucial to recognize that these policies not only affect costs and rewards but also influence system reliability. One of the main challenges is optimizing the vacation duration: longer vacation periods increase production/revenue losses and costs incurred due to failed machines in the queue, while shorter vacation periods lead to increased idleness and system costs. A well-designed vacation policy for the repairperson effectively balances the system according to the optimal distribution. In Shekhar et al. [13], a comparative study of different vacation policies on the reliability characteristics of machining systems was presented. The incorporation of maintenance policies under Bernoulli vacation has been considered in the field of queuing theory [14], Wu and Lan [15], Rathore and Shrivastava [16]). A redundant repairable series system is analyzed in Gao et al. [17]. The authors propose a system consisting of a fixed number of units, each equipped with its own standby counterpart. Various types of failures are considered, including individual unit failures, switching failures involving standby units, and simultaneous failures of all units. A vacation policy for the repairperson is introduced: if all units remain operational for a certain period, the repairperson enters a vacation mode, returning only upon the occurrence of a failure. The analysis is conducted using a Markov process and a second-order linear nonhomogeneous difference equation with variable coefficients. An optimization problem is formulated based on the expected total cost per unit time, subject to constraints, and solved using the Sequential Least Squares Programming (SLSQP) algorithm.

An important difficulty encountered when modeling complex multi-state systems is the presence of intractable expressions in both the modeling process and the performance functions. This issue complicates algorithmization and the interpretation of results. An important aspect to consider is the development of methodologies that make both the modeling process and the reliability metrics of complex systems more accessible. Simplifying these approaches can facilitate their application in real-world scenarios while maintaining a high level of accuracy and relevance. One option that facilitates the modeling of these systems is the use of phase-type (PH) distributions and Markovian Arrival Processes (MAP). Phase-type distributions can be defined as the time until absorption in an absorbing Markov chain. This class of distributions was introduced by Neuts [18] and analyzed in detail in Neuts [19]. These distributions possess excellent properties, with one particularly noteworthy property being their density in the set of non-negative probability distributions [20]. This property implies that any non-negative probability distribution can be approximated arbitrarily closely by a phase-type distribution. This feature allows us to consider that the probability distributions embedded in the model are general. On the other hand, Markovian Arrival Processes, introduced by Neuts [21], generalize counting processes. Both structures allow for the algorithmic modeling of complex systems, facilitating the interpretation of results and simplifying computational implementation. For example, Yu et al. [22] proposed a repair model for a device based on a phase-type geometric process for both failure and repair times, incorporating multiple server vacations. Eryilmaz [23] introduced a phase-type modeling approach for dynamically assessing non-repairable multi-state systems with various performance levels. Kim and Kim [24] addressed redundancy allocation problems (RAP), considering different redundancy strategies and the implications of imperfect switching in standby redundant systems using PH distributions. Cui and Wu [25] introduced two extended PH models tailored to multi-state competing risk systems. Recently, Li et al. [26] proposed efficient techniques leveraging PH distributions to accurately assess and predict the reliability of systems with time-varying characteristics. Sun and Vatn [27] developed a maintenance model integrating condition-based inspections and maintenance delays using PH distributions. Pérez-Ocón et al. [28] utilized PH distributions to model an M-system incorporating geometric processes applied to a G-out-of-M system. Ruiz-Castro et al. [29] modeled multi-component systems using algorithmic methods based on PH distributions. Significant studies on modeling resistive memories using PH distributions have been conducted by Ruiz-Castro et al. [30]. They analyzed the behaviour of complex systems through macro-states using PH distributions in the context of random telegraph noise in resistive memories. Albrecher and Bladt [31] and Ruiz-Castro et al. [32] have introduced non-homogeneity into phase-type distributions, paving the way for new and more complex models in the field of reliability. The first work introduces inhomogeneous phase-type distributions to analyze heavy-tailed distributions and the second study constructed step-wise non-homogeneous PH distributions to model complex systems applied to resistive memories with a new set of electron devices for nonvolatile memory circuits.

Sophisticated complex systems incorporating preventive maintenance and multiple vacation policies using Marked Markovian Arrival Processes (MMAPs) have also been developed. Ruiz-Castro [33] analyzed a complex multi-state redundant system with preventive maintenance, multiple events, and a multiple vacation policy using MMAPs. Ruiz-Castro and Acal [34] modeled two complex reliability systems with and without preventive maintenance, applying a multiple vacation policy for the repairperson.

In this work, a complex multi-state system subject to various events is modeled by incorporating a preventive maintenance strategy based on degradation levels, along with a Bernoulli multiple vacation policy for the repairperson. This approach aims to capture more realistic operational dynamics and improve system reliability through strategic workforce management and maintenance scheduling. The proposed model reflects real-world systems. In civil engineering, urban water distribution networks exhibit similar behaviour. In industrial engineering, automated production lines involve multiple interconnected machines subject to failures, periodic technician absence, preventive maintenance strategies, and external shocks like supply chain interruptions. In the field of computer engineering, distributed data centers show analogous patterns, where servers of varying criticality undergo recoverable or permanent failures, technician availability is uncertain due to multitasking or leave, and external events such as cyber-attacks or power outages introduce random shocks to the system.

The internal behaviour of the considered system is partitioned into an indeterminate number of functioning/degradation levels, which may experience either repairable or non-repairable failures. The system is also subject to external shocks. An external shock may cause internal deterioration of the performance level, potentially leading to repairable or non-repairable failures, total irreparable failure, or external accumulated damage. Once this damage reaches a certain threshold, the unit is restored. To optimize the system, the repairperson operates under a Bernoulli multiple vacation policy. Preventive maintenance is incorporated, ensuring that when the repairperson observes a critical degradation level upon returning, the unit is sent to the repair channel. The system is modeled using PH distributions and MMAPs, enabling algorithmic and matrix-based expressions for the process and the measures obtained. The modeling is carried out in both continuous and discrete time, analyzing both transient and stationary cases. A key aspect of this study is optimizing the vacation policy by analyzing system costs and availability.

This work extends previous research in the field in the following ways: 1) It provides an algorithmic matrix-based formulation, constructing a novel MMAP, for a complex system that evolves through a specified number of degradation levels, each comprising multiple operational phases. The proposed MMAP model enables the modeling and analysis of a complex multi-state reliability system in a mathematically rigorous manner through matrix-block structures. The model allows for the computation of performance measures and the incorporation of cost/benefit factors algorithmically, yielding interpretable results. 2) It incorporates a critical operational level that triggers preventive maintenance. 3) The internal behaviour of the system is subject to both repairable and non-repairable failures, introducing two absorbing states into the corresponding phase-type distribution. 4) External shocks with multiple possible consequences are introduced. 5) Costs and benefits associated with the system are incorporated algorithmically, depending on various factors. 6) The model is constructed using an algorithmic-matrix approach, along with associated performance measures, both in transient and steady-state regimes. 7) In addition to the continuous-time case, the discrete-time case is modeled, increasing the complexity due to the possibility of multiple events occurring simultaneously. 8) An optimization policy is developed for the vacation time and the repairperson's decision-making, considering the system's degradation level, cost/benefit structure, and operational status.

Therefore, this work develops a new model and the corresponding associated measures, combining a novel MMAP and phase-type distributions, integrating matrix analysis for complex systems with preventive and corrective maintenance, and a novel, realistic vacation policy. To obtain the optimal system, a novel methodology in reliability is introduced. All of this is developed in both continuous and discrete time, as well as in transient and steady-state regimes.

The structure of the work is as follows. Section 2 describes the system and the vacation policy. The model is detailed in Section 3 for both the continuous and discrete cases. Stationary distributions are calculated using analytical and matrix-based methods for both cases in Section 4. Multiple metrics, such as availability, reliability, and the mean number of events, are provided in Section 5. In Section 6, costs and rewards are incorporated into the model in a matrix-based framework. Optimization is presented in Section 7. Finally, in Section 8, a numerical example is developed to highlight the potential of the proposed methodology. Conclusions are provided in Section 9.

## 2. System behaviour and vacation policy

We assume a system with a multi-state unit which has an internal behaviour with an indeterminate performance. A general number of performance/degradation levels are considered, $K$. In turn, the number of states for the $k$-th level is $n_k$. The unit may experience repairable and non-repairable failures from all levels of degradation. The unit may undergo external impacts, which in turn may cause non-repairable failures, modifications to the internal structure (even leading to repairable and non-repairable failures), and cumulative external damage (which can lead to non-repairable failures when it reaches a certain threshold).

The repair facility is composed of a repairperson, who may take Bernoulli vacation periods depending on the degradation level of the unit. The new vacation policy is outlined as follows. Initially, the repairperson is on vacation. When the repairperson returns, if the system is below the $K$-st degradation level (critical level), a new random vacation period will begin for the repairperson with probability $p_k$, for $k = 1, 2, \ldots, K - 1$. Otherwise, the repairperson stays at the workplace waiting for a possible repairable failure.

If the repairperson is on vacation and returns to find the unit at the last level of degradation, critical level, the unit undergoes preventive maintenance. However, if the repairperson is on vacation and a repairable failure occurs, the unit will remain in the "repairable failure" macro-state until the repairperson returns and starts the corrective repair. Similarly, if the repairperson is on vacation and the unit undergoes a non-repairable failure, it will be replaced with an identical unit when the repairperson returns. Otherwise, if the repairperson is at the workplace and a repairable or non-repairable failure occurs, the repair will start immediately or the unit will be replaced in a negligible amount of time, respectively. All random times considered in the model pass through various states until the event occurs.

A diagram of the Bernoulli vacation policy is given in Fig. 1.

## 3. The model

The system has been modeled in both continuous and discrete time, considering a Markovian arrival process with marked arrivals. Both models will be presented in parallel, taking into account that the simultaneous events that may occur are not the same. Special attention will be given to the continuous time. The assumptions are presented in Section 3.1. Section 3.2 introduces the state space, while the MMAP is developed in Section 3.3 for the continuous case and in Section 3.4 for the discrete case.

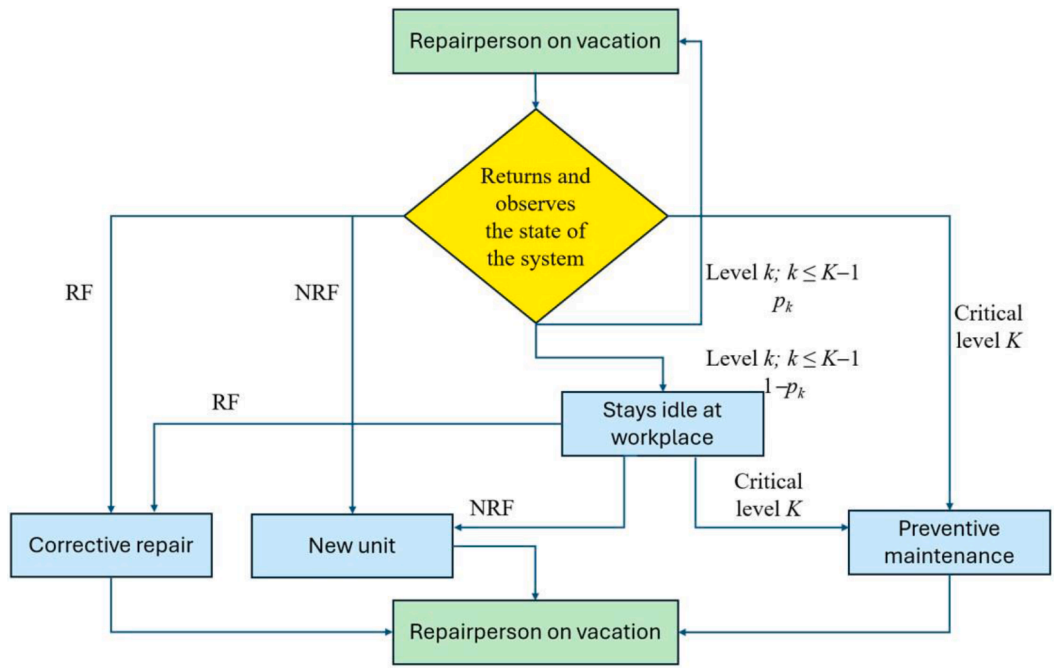

**Fig. 1.** Diagram of the Bernoulli vacation policy.

### 3.1. Assumptions

The assumptions are introduced for the continuous and discrete cases.

#### 3.1.1. The continuous case
The system, that evolves in continuous time, follows the following assumptions:

**Assumption 1.** The number of internal performance/degradation levels is $K$.

**Assumption 2.** The number of phases in degradation in the $k$-th level is $n_k$ for $k = 1, 2, \ldots, K$.

**Assumption 3.** The internal performance time follows a PH distribution with $(\boldsymbol{\alpha}, \mathbf{T})$ with order $m = n_1 + n_2 + \cdots + n_K$. The PH representation is composed of a matrix blocks according to the levels. The matrix $\mathbf{T}$ is given by

$$\mathbf{T} = \begin{pmatrix} \mathbf{T}_{11} & \mathbf{T}_{12} & \cdots & \mathbf{T}_{1K} \\ \mathbf{T}_{21} & \mathbf{T}_{22} & \cdots & \mathbf{T}_{2K} \\ \vdots & \vdots & \ddots & \vdots \\ \mathbf{T}_{K1} & \mathbf{T}_{K2} & \cdots & \mathbf{T}_{KK} \end{pmatrix}.$$

The order of $\mathbf{T}_{ii}$ is $n_i$ para $i = 1, 2, \ldots, K$. The column vector $\mathbf{T}^0 = -\mathbf{T}\mathbf{e}$ contains the failures rates from the different operational rates. In this paper $\mathbf{e}$ is a column vector of ones with appropriate order and $\mathbf{A}^0 = -\mathbf{A}\mathbf{e}$ for any matrix $\mathbf{A}$. The column vector is expressed as $\mathbf{T}^0 = \mathbf{T}_r^0 + \mathbf{T}_{nr}^0$, where $\mathbf{T}_r^0$ and $\mathbf{T}_{nr}^0$ are column vectors which contain the internal repairable and non-repairable transition rates, respectively.

**Assumption 4.** The external shock is modelled through a PH renewal process where the time between two consecutive shocks is PH distributed with representation $(\boldsymbol{\gamma}, L)$ with order $t$. The vector $\mathbf{L}^0 = -\mathbf{L}\mathbf{e}$ contains the transitions intensities up to external shock time. External shocks occur independently from the condition of the system.

**Assumption 5.** The occurrence of external shocks can produce changes in the internal performance. In this case, $\mathbf{W}$ represents the probability matrix of modification of internal performance after shock. The $(i, j)$ element of this matrix contains the probability of transitioning from phase $i$ to phase $j$ following a shock affecting the unit's internal performance. The vector $\mathbf{W}_r^0$ and $\mathbf{W}_{nr}^0$ contains the probabilities of the transition to internal repairable and the probabilities of the transition to internal non-repairable after shock, respectively.

**Assumption 6.** An external shock can provoke a total failure of the unit after it with a probability $\omega^0$.

**Assumption 7.** After a shock, cumulative damage is produced. This damage progresses through distinct phases. The probability matrix $\mathbf{C}$ represents the probabilities of transitioning between cumulative damage phases. When a new unit is introduced, the initial cumulative damage depends on the phases given by $\omega$. Typically, the first phase of cumulative damage corresponds to a state with no damage.

**Assumption 8.** The corrective repair time follows a PH distribution with representation $(\boldsymbol{\beta}^1, \mathbf{S}_1)$, being $\mathbf{S}_1$ a matrix of order $m_1$.

**Assumption 9.** The preventive maintenance time follows a PH distribution with representation $(\boldsymbol{\beta}^2, \mathbf{S}_2)$, being $\mathbf{S}_2$ a matrix of order $m_2$.

**Assumption 10.** Initially, the repairperson is on vacation. The vacation time follows a PH distribution with representation $(\boldsymbol{\upsilon}, \mathbf{V})$, being $\mathbf{V}$ a matrix of order $v$.

**Assumption 11.** The probability of begin a new vacation period after returning with internal degradation level $k$ is $p_k$, for $k = 1, 2, \ldots, K-1$. If the repairperson observes the degradation level $K$ (critical level) upon returning from the vacation period, preventive maintenance is initiated.

**Assumption 12.** After completing preventive maintenance, a corrective repair, or replacing the unit, the repairperson begins a new vacation period.

#### 3.1.2. The discrete case
For the discrete case, all assumptions are similar to those of the continuous case, with the following considerations.

The phase-type distributions $(\boldsymbol{\alpha},\mathbf{T}), (\boldsymbol{\gamma},L), (\boldsymbol{v},\mathbf{V}), (\boldsymbol{\beta}^1,\mathbf{S}_1), (\boldsymbol{\beta}^2,\mathbf{S}_2)$ are discrete, and therefore, the matrices $\mathbf{T, L, V, S}_1$ and $\mathbf{S}_2$ contain transition probabilities instead of transition rates.

Additionally, in the discrete case, it holds that $\mathbf{A}^0 = \mathbf{e} - \mathbf{Ae}$ for any matrix $\mathbf{A}$. Consequently, the matrices $\mathbf{A}^0$ contain the absorption probabilities for each event based on the various phases.

### 3.2. State-Space

The system is governed by a vector Markov process with a state-space composed of the following macro-states (similar for the continuous and discrete time)

$$S = \{O^v, O^{nv}, RF, NRF, CR, PM\},$$

where the macro-states indicate

$O^Y$ = the unit is working, and the repairperson is on vacation ($Y = v$) or at the workplace ($Y=nv$).

$RF$ = the unit underwent a repairable failure, and the repairperson is on vacation.

$NRF$ = the unit underwent a non-repairable failure, and the repairperson is on vacation.

$CR$ = the unit is in corrective repair with the repairperson

$PM$ = the unit is in preventive maintenance with the repairperson

The phases of the macro-states for the state-space are the following,

$E_1 = O^v = \{(i,j,u,s) | i=1,\ldots,m; j=1,\ldots,t; u=1,\ldots,d; s=1,\ldots,v\}.$

The system is working in internal phase $i$, the external shock time is in phase $j$, the external damage of the unit is in state $u$ and finally the vacation time is in phase $s$,

$E_2 = O^{nv} = \{(i,j,u) | i=1,\ldots,m-n_K; j=1,\ldots,t; u=1,\ldots,d\}.$

The system is working in internal phase $i$, the external shock time is in phase $j$ and the external damage of the unit is in state $u$,

$E_3 = RF = \{(j,s) | j=1,\ldots,t; s=1,\ldots,v\}.$

The system has undergone a repairable failure, the external shock time is in phase $j$ and the vacation time is in phase $s$,

$E_4 = NRF = \{(j,s) | j=1,\ldots,t; s=1,\ldots,v\}.$

The system has underwent a non-repairable failure, the external shock time is in phase $j$ and the vacation time is in phase $s$,

$E_5 = CR = \{(j,r) | j=1,\ldots,t; r=1,\ldots,m_1\}.$

The system is in corrective repair in corrective repair phase $r$ and the external shock time is in phase $j$,

$E_6 = PM = \{(j,r) | j=1,\ldots,t; r=1,\ldots,m_2\}.$

The system is in preventive maintenance phase $r$ and the external shock time is in phase $j$.

### 3.3. The continuous MMAP

In the continuous case, the system may undergo the following types of events which are denoted as,

*RF*: The unit is operational but experiences a repairable failure (the repairperson is on vacation). At that moment, the system ceases to function and waits for the repairperson's availability.

*RF+CR*: The unit is working, it undergoes a repairable failure, and it begins corrective repair because the repairperson was in the workplace.

*NRF*: The unit is operational but experiences a non-repairable failure (the repairperson is on vacation). At that moment, the system stops functioning and waits for the repairperson's presence to be replaced.

*NRF+NU*: The unit is working, and it undergoes a non-repairable failure, and it is replaced because the repairperson was in the workplace.

*PM*: The unit is working, and it crosses to the last internal damage level (critical level) and it begins preventive maintenance (the repairperson was in the workplace).

*R*: The only event is that the repairperson returns and stays at the workplace.

*R+NVP*: The repairperson returns from vacation and begins a new vacation period.

*R+PM*: The repairperson returns from vacation, observes major damage (critical level), and begins preventive maintenance.

*R+CR*: The repairperson returns from vacation, observes previous repairable failure and begins corrective repair.

*R+NU*: The repairperson returns from vacation, observes previous non-repairable failure and, it is replaced by a new and identical one. A new vacation period begins.

The novel MMAP linked to the system has been constructed based on the various events discussed earlier. The representation is

$$\{\mathbf{Q}^0, \mathbf{Q}^{RF}, \mathbf{Q}^{NRF}, \mathbf{Q}^R, \mathbf{Q}^{PM}, \mathbf{Q}^{RF+CR}, \mathbf{Q}^{NRF+NU}, \mathbf{Q}^{R+CR}, \mathbf{Q}^{R+NU}, \mathbf{Q}^{R+PM}, \mathbf{Q}^{R+NVP}\},$$

where $\mathbf{Q}^Y$ contains the transition intensities for the event $Y$ and $\mathbf{Q}^0$ with transitions no events.

The new matrix functions constructed for modeling the various events will be described below.

#### 3.3.1. Auxiliary Matrices

To construct the matrix functions for subsequent modeling, various auxiliary matrices are developed. These matrices allow consideration of the different internal operational levels of the system during modeling.

- Matrix $\mathbf{U}_k$. The elements of auxiliary matrices $\mathbf{U}_k$ with order $m \times m$ from level $k$ to $k$ are,

$$\mathbf{U}_k(i,j) = \begin{cases} 1; & i=j; \sum_{h=1}^{k-1} n_h + 1 \leq i,j \leq \sum_{h=1}^{k} n_h \\ 0; & \text{otherwise}. \end{cases}$$

This matrix enables consideration of transitions solely from states of level $k$. In other words, if, for example, the internal transition matrix $\mathbf{T}$ is multiplied on the left by $\mathbf{U}_k$, the result is a matrix of the same order whose only nonzero elements correspond to those of $\mathbf{T}$ in the rows of level $k$; thus, only transitions from level $k$ are considered.

- Matrix $\mathbf{U}_{1,\ldots,K-1}$. Matrix with order $(m-n_k) \times m$. From levels 1 to $K$-1 where the beginning is restricted to the first $K$-1 states. This is

$$\mathbf{U}_{1,\ldots,K-1} = \left(\mathbf{I}_{n_1+n_2+\cdots+n_{K-1}} | \mathbf{0}_{n_1+n_2+\cdots+n_{K-1} \times n_K}\right),$$

where, throughout the paper, the matrix $\mathbf{I}$ is the identity matrix with appropriate order.

When the internal behaviour matrix is multiplied on the left by this matrix, the result is another matrix containing only transitions from the phases of the first $K$-1 levels to any level. This matrix will be used primarily when transitions are to be made from any non-critical level.

- Matrix $\mathbf{U}'_{1,\ldots,K-1}$. Matrix with order $m \times (m-n_K)$. From levels 1 to $K$-1 where the ending is restricted to the first $K$-1 states. This is the transpose of $\mathbf{U}_{1,\ldots,K-1}$.

This matrix will be used in the modeling where the focus is placed after a transition. That is, if the internal behaviour matrix is multiplied on the right by the defined matrix, the result is a matrix containing transitions from any level but only to non-critical levels (the rest are zeros).

- Clearly, $\mathbf{U}_{1,\ldots,K} = \mathbf{U}'_{1,\ldots,K} = \mathbf{I}.$

The construction of these new matrices is essential in the algorithmization of the system.

### 3.3.2. Repairable failure function ($H_{RF}$)

The unit is working, and it undergoes a repairable failure. This can occur at a certain operational level, as determined by the general matrix **U**. The matrix **U** indicates from which levels internal transitions are possible. For this event, the following scenarios are possible for the unit system:

- An internal failure occurs with no external shock, $\mathbf{UT}_r^0 \otimes \mathbf{I} \otimes \mathbf{e}$.
- An external shock occurs but does not result in total non-repairable failure $(1 - \omega^0)$. This shock induces a repairable failure by altering the degradation level. The external damage threshold is not reached (**Ce**). Then $\mathbf{UW}_r^0 \otimes \mathbf{L}^0\gamma(1-\omega^0) \otimes \mathbf{Ce}$.

Then, the matrix function for repairable failure is given by

$$\mathbf{H}_{RF}(\mathbf{U}) = \mathbf{UT}_r^0 \otimes \mathbf{I} \otimes \mathbf{e} + \mathbf{UW}_r^0 \otimes \mathbf{L}^0\gamma(1-\omega^0) \otimes \mathbf{Ce}. \quad (1)$$

### 3.3.3. Non-repairable failure function ($H_{NRF}$)

The unit is working, and a non-repairable failure occurs. This can also occur at a certain operational level, determined by the general matrix **U**. Additionally, the function depends on vectors **R** and **A** based on what happens after the non-repairable failure to the main unit (internal behaviour and external damages, respectively). This is the case because, after the non-repairable failure, the unit may either remain broken waiting for the repairperson or be replaced if the repairperson is present in the workplace. The following scenarios are possible:

- an internal non-repairable failure occurs with no external shock, $\mathbf{UT}_{nr}^0 \mathbf{R} \otimes \mathbf{I} \otimes \mathbf{eA}$.
- an external shock occurs but does not provoke total failure $(1 - \omega^0)$. This shock provokes a modification in the internal performance and this modification causes a non-repairable internal failure, this external shock changes the degradation level but the threshold is not reached, $\mathbf{UW}_{nr}^0 \mathbf{R} \otimes \mathbf{L}^0\gamma(1-\omega^0) \otimes \mathbf{CeA}$.
- an external shock provokes a non-repairable failure, $\mathbf{UeR} \otimes \mathbf{L}^0\gamma\omega^0 \otimes \mathbf{eA}$.
- an external shock occurs, but does not provoke total failure, but this shock causes a change in the external degradation level provoking a non-repairable failure, this failure is not produced by internal causes, $\mathbf{UeR} \otimes \mathbf{L}^0\gamma(1-\omega^0) \otimes \mathbf{C}^0\mathbf{A}$.

Then, the matrix is given by

$$\mathbf{H}_{NRF}(\mathbf{U},\mathbf{R},\mathbf{A}) = \mathbf{UT}_{nr}^0\mathbf{R} \otimes \mathbf{I} \otimes \mathbf{eA} + \mathbf{UW}_{nr}^0\mathbf{R} \otimes \mathbf{L}^0\gamma(1-\omega^0) \otimes \mathbf{CeA}$$
$$+ \mathbf{UeR} \otimes \mathbf{L}^0\gamma\omega^0 \otimes \mathbf{eA} + \mathbf{UeR} \otimes \mathbf{L}^0\gamma(1-\omega^0) \otimes \mathbf{C}^0\mathbf{A}. \quad (2)$$

### 3.3.4. No events at certain time function ($H_O$)

We assume the unit does not experience any event described in Section 3.3 at this time and continues working. This function depends on the internal system state before the transition (**U**), the internal state after the transition (**R**), and the external degradation state following the transition (**A**). This occurs due to various situations:

- The internal performance continues in the same phase or changes to another, equally operational state. There is no external shock, $\mathbf{UTR} \otimes \mathbf{I} \otimes \mathbf{A}$.
- The repairperson is in the workplace, there is not an external shock, $\mathbf{UR} \otimes \mathbf{L} \otimes \mathbf{A}$.
- The repairperson is in the workplace, there is an external shock that does not produce a non-repairable failure, but the unit continues working, this shock does not provoke changes in the degradation level, $\mathbf{UWR} \otimes \mathbf{L}^0\gamma(1-\omega^0) \otimes \mathbf{CA}$.

Then the matrix that governs this is given by

$$\mathbf{H}_O(\mathbf{U},\mathbf{R},\mathbf{A}) = \mathbf{UTR} \otimes \mathbf{I} \otimes \mathbf{A} + \mathbf{UR} \otimes \mathbf{L} \otimes \mathbf{A} + \mathbf{UWR} \otimes \mathbf{L}^0\gamma(1-\omega^0) \otimes \mathbf{CA}. \quad (3)$$

Next, the matrix $\mathbf{Q}^{PM}$ is described in the continuous case. The rest are given in the Appendix A.

### 3.3.5. The matrix $\mathbf{Q}^{PM}$

The matrix $\mathbf{Q}^{PM}$ governs the transition when the unit is working, it crosses to the $K$-th degradation level (critical level) and it begins a preventive maintenance because the repairperson was in the workplace.

$$\mathbf{Q}^{PM} = \begin{array}{c} O^v \\ O^{nv} \\ RF \\ NRF \\ CR \\ PM \end{array} \begin{pmatrix} 0 & 0 & 0 & 0 & 0 & 0 \\ 0 & 0 & 0 & 0 & 0 & \mathbf{Q}^{PM}_{O^{nv},PM} \\ 0 & 0 & 0 & 0 & 0 & 0 \\ 0 & 0 & 0 & 0 & 0 & 0 \\ 0 & 0 & 0 & 0 & 0 & 0 \\ 0 & 0 & 0 & 0 & 0 & 0 \end{pmatrix},$$

where

$$\mathbf{Q}^{PM}_{O^{nv},PM} = \mathbf{H}_O(\mathbf{U}_{1,\ldots,K-1}, \mathbf{U}_K\mathbf{e}, \mathbf{e}) \otimes \boldsymbol{\beta}^2$$
$$= \left[\mathbf{U}_{1,\ldots,K-1}\mathbf{TU}_K\mathbf{e} \otimes \mathbf{I} \otimes \mathbf{e} + \mathbf{U}_{1,\ldots,K-1}\mathbf{WU}_K\mathbf{e} \otimes \mathbf{L}^0\gamma(1-\omega^0) \otimes \mathbf{Ce}\right] \otimes \boldsymbol{\beta}^2,$$

being $\mathbf{H}_O(\cdot,\cdot,\cdot)$ the matrix function given in (3).

Two possibilities arise,

- the system was at a level below $K$ ($\mathbf{U}_{1,\ldots,K-1}$), and the only transition is to the critical level $K$ ($\mathbf{TU}_K$), neither the timing of the external shock nor the wear caused by external shocks changes and the preventive maintenance begins with an initial distribution, $\boldsymbol{\beta}^2$. That is $\mathbf{U}_{1,\ldots,K-1}\mathbf{TU}_K\mathbf{e} \otimes \mathbf{I} \otimes \mathbf{e} \otimes \boldsymbol{\beta}^2$.
- an external shock occurs without a non-repairable failure ($\mathbf{L}^0\gamma(1-\omega^0) \otimes \mathbf{Ce}$), and from a level below critical, the system transitions to critical $K$ due to the shock, $\mathbf{U}_{1,\ldots,K-1}\mathbf{WU}_K\mathbf{e}$. Preventive maintenance immediately begins with initial distribution $\boldsymbol{\beta}^2$.

### 3.4. The discrete MMAP

In the discrete case, the following events could occur, considering that simultaneous events can occur at the same time.

*RF*: the unit is working, and it undergoes a repairable failure (the repairperson is on vacation).

*NRF*: the unit is working, and it undergoes a non-repairable failure (the repairperson is on vacation).

*PM*: the unit is working, it crosses to the critical internal damage level and it begins preventive maintenance (the repairperson was in the workplace).

*R*: the only event is that the repairperson returns and stays at the workplace.

*NRF+NU*: the unit is working, and it undergoes a non-repairable failure, and it is replaced because the repairperson was in the workplace.

*RF+CR*: the unit is working, it undergoes a repairable failure, and it begins corrective repair because the repairperson was in the workplace.

*R+PM*: the repairperson returns from vacation, observes critical level $K$, and begins preventive maintenance.

*R+CR*: the repairperson returns from vacation, observes previous repairable failure and, begins corrective repair.

*R+NU*: the repairperson returns from vacation, observes previous non-repairable failure and, it is replaced by a new and identical one.

*R+NVP*: the repairperson returns from vacation and begins a new vacation period.

*R+RF+CR*: the unit is working, it undergoes a repairable failure, and at the same time the repairperson returns from vacation and begins corrective repair.

*R+NRF+NU*: the unit is working, undergoes a non-repairable failure and at the same time the repairperson returns from vacation and the unit is renewed.

The representation of the discrete MMAP is

$$\{\mathbf{D}^O, \mathbf{D}^{RF}, \mathbf{D}^{NRF}, \mathbf{D}^{CR}, \mathbf{D}^{PM}, \mathbf{D}^{RF+CR}, \mathbf{D}^{NRF+NU}, \mathbf{D}^{R+CR}, \mathbf{D}^{R+NU}, \mathbf{D}^{R+PM}, \mathbf{D}^{R+RF+CR}, \mathbf{D}^{R+NRF+NU}, \mathbf{D}^{R+NVP}\}.$$

### 3.4.1. The matrix $\mathbf{D}^{PM}$ for the discrete case

Analogously to the continuous case, the $\mathbf{H}_{RF}$, $\mathbf{H}_{NRF}$, and $\mathbf{H}_O$ functions have been constructed for the discrete case. In this context, it is important to consider that multiple events may occur simultaneously, which makes the modeling process somewhat more complex. These functions are described in Appendix B1.

Having defined the matrix functions $\mathbf{H}$ for the discrete case, the general structure of the matrix $\mathbf{D}^{PM}$ is similar to that of the continuous case. Thus,

$$\mathbf{D}^{PM} = \begin{matrix} O^v \\ O^{nv} \\ RF \\ NRF \\ CR \\ PM \end{matrix} \begin{pmatrix} 0 & 0 & 0 & 0 & 0 & 0 \\ 0 & 0 & 0 & 0 & 0 & \mathbf{D}^{PM}_{O^{nv},PM} \\ 0 & 0 & 0 & 0 & 0 & 0 \\ 0 & 0 & 0 & 0 & 0 & 0 \\ 0 & 0 & 0 & 0 & 0 & 0 \\ 0 & 0 & 0 & 0 & 0 & 0 \end{pmatrix},$$

where

$$\mathbf{D}^{PM}_{O^{nv},PM} = \mathbf{U}_{1,\ldots K-1}\mathbf{TU}_K \mathbf{e} \otimes \mathbf{L} \otimes \mathbf{e} \otimes \boldsymbol{\beta}^2 + \mathbf{U}_{1,\ldots K-1}\mathbf{TWU}_K \mathbf{e} \otimes \mathbf{L}^0\boldsymbol{\gamma}(1-\omega^0)$$
$$\otimes \mathbf{Ce} \otimes \boldsymbol{\beta}^2.$$

The system is in a non-critical level ($\mathbf{U}_{1,\ldots K-1}$) and transitions to a critical level either without a shock ($\mathbf{TU}_K$), or after a shock ($\mathbf{L}^0\boldsymbol{\gamma}$) that could worsen the internal level ($\mathbf{TWU}_K$), always without experiencing a non-repairable failure, whether due to an extreme shock ($1-\omega^0$) or by accumulation ($\mathbf{Ce}$). The preventive maintenance begins with initial probability $\boldsymbol{\beta}^2$.

The rest of matrix-blocks are given in Appendix B2.

## 4. The transient and stationary distribution

In this section the transient and stationary distribution are calculated by using matrix-analytic methods.

### 4.1. The transient distribution

The MMAP described in Section 3.3 is governed by a Markov process with generator

$$\mathbf{Q} = \mathbf{Q}^0 + \mathbf{Q}^{RF} + \mathbf{Q}^{NRF} + \mathbf{Q}^R + \mathbf{Q}^{PM} + \mathbf{Q}^{RF+CR} + \mathbf{Q}^{NRF+NU} + \mathbf{Q}^{R+CR} + \mathbf{Q}^{R+NU} + \mathbf{Q}^{R+PM} + \mathbf{Q}^{R+NVP}.$$

Initially, the system is brand new, and the repairperson starts a vacation. Therefore, the initial distribution for the system is

$$\boldsymbol{\theta} = (\boldsymbol{\alpha} \otimes \boldsymbol{\pi}_L^c \otimes \boldsymbol{\omega} \otimes \boldsymbol{\upsilon}, \mathbf{0}),$$

where $\boldsymbol{\pi}_L^c$ is the stationary distribution for the external shock time. It is for the continuous case equal to $\boldsymbol{\pi}_L^c = (1, \mathbf{0})[\mathbf{e}|(\mathbf{L}+\mathbf{L}^0\boldsymbol{\gamma})^*]^{-1}$, and for the discrete case $\boldsymbol{\pi}_L^d = (1, \mathbf{0})[\mathbf{e}|(\mathbf{I}-\mathbf{L}-\mathbf{L}^0\boldsymbol{\gamma})^*]^{-1}$, where $\mathbf{M}^*$ is the matrix $\mathbf{M}$ without the first column.

The transient distribution, for the continuous case, representing the probability of being in state *i* at time *t*, is given by the *i*-th element of

$$\mathbf{p}(t) = \boldsymbol{\theta}e^{\mathbf{Q}t}, \tag{4}$$

and for the discrete case this probability is

$$\mathbf{p}^v = \boldsymbol{\theta}\mathbf{D}^v, \tag{5}$$

being $\mathbf{D}$ the transition probability matrix, achieved from the corresponding MMAP for the discrete case given in Section 3.4. That is,

$$\mathbf{D} = \mathbf{D}^O + \mathbf{D}^{RF} + \mathbf{D}^{NRF} + \mathbf{D}^{CR} + \mathbf{D}^{PM} + \mathbf{D}^{RF+CR} + \mathbf{D}^{NRF+NU} + \mathbf{D}^{R+CR} + \mathbf{D}^{R+NU} + \mathbf{D}^{R+PM} + \mathbf{D}^{R+RF+CR} + \mathbf{D}^{R+NRF+NU} + \mathbf{D}^{R+NVP}.$$

These vectors can be divided into macro-states $E_s$, $s = 1,\ldots, 6$, with

$$\mathbf{p}_{E_s}(t) = (\boldsymbol{\theta}e^{\mathbf{Q}t})_{E_s}, \tag{6}$$

for the continuous case, and for the discrete case

$$\mathbf{p}^v_{E_s} = (\boldsymbol{\theta}\mathbf{D}^v)_{E_s}. \tag{7}$$

### 4.2. The stationary distribution. Continuous case

The stationary distribution has been derived using matrix-algorithmic methods. The stationary distribution, denoted by $\boldsymbol{\pi}$, is segmented according to the macro-state space $S$, $\boldsymbol{\pi} = \{\boldsymbol{\pi}_1, \boldsymbol{\pi}_2, \boldsymbol{\pi}_3, \boldsymbol{\pi}_4, \boldsymbol{\pi}_5, \boldsymbol{\pi}_6\}$. This vector satisfies the balance matrix equation jointly the normalization condition,

$$\boldsymbol{\pi}\mathbf{Q} = \mathbf{0}; \boldsymbol{\pi}\mathbf{e} = 1$$

The transition intensities matrix can be expressed as,

$$\mathbf{Q} = \begin{matrix} O^v \\ O^{nv} \\ RF \\ NRF \\ CR \\ PM \end{matrix} \begin{pmatrix} \mathbf{Q}_{11} & \mathbf{Q}_{12} & \mathbf{Q}_{13} & \mathbf{Q}_{14} & 0 & \mathbf{Q}_{16} \\ \mathbf{Q}_{21} & \mathbf{Q}_{22} & 0 & 0 & \mathbf{Q}_{25} & \mathbf{Q}_{26} \\ 0 & 0 & \mathbf{Q}_{33} & 0 & \mathbf{Q}_{35} & 0 \\ \mathbf{Q}_{41} & 0 & 0 & \mathbf{Q}_{44} & 0 & 0 \\ \mathbf{Q}_{51} & 0 & 0 & 0 & \mathbf{Q}_{55} & 0 \\ \mathbf{Q}_{61} & 0 & 0 & 0 & 0 & \mathbf{Q}_{66} \end{pmatrix},$$

where,

$$\mathbf{Q}_{11} = \mathbf{Q}^{R+NVP}_{O^v,O^v} + \mathbf{Q}^O_{O^v,O^v}; \mathbf{Q}_{12} = \mathbf{Q}^R_{O^v,O^{nv}}, \mathbf{Q}_{13} = \mathbf{Q}^{RF}_{O^v,RF}, \mathbf{Q}_{14} = \mathbf{Q}^{NRF}_{O^v,NRF}, \mathbf{Q}_{16} = \mathbf{Q}^{R+PM}_{O^v,PM},$$

$$\mathbf{Q}_{21} = \mathbf{Q}^{NRF+NU}_{O^{nv},O^v}, \mathbf{Q}_{22} = \mathbf{Q}^O_{O^{nv},O^{nv}}, \mathbf{Q}_{25} = \mathbf{Q}^{RF+CR}_{O^{nv},CR}, \mathbf{Q}_{26} = \mathbf{Q}^{PM}_{O^{nv},PM}, \mathbf{Q}_{33} = \mathbf{Q}^O_{RF,RF},$$

$$\mathbf{Q}_{35} = \mathbf{Q}^{R+CR}_{RF,CR}, \mathbf{Q}_{41} = \mathbf{Q}^{R+NU}_{NRF,O^v}, \mathbf{Q}_{44} = \mathbf{Q}^O_{NRF,NRF}, \mathbf{Q}_{51} = \mathbf{Q}^O_{CR,O^v}, \mathbf{Q}_{55} = \mathbf{Q}^O_{CR,CR},$$

$$\mathbf{Q}_{61} = \mathbf{Q}^O_{PM,O^v}, \mathbf{Q}_{66} = \mathbf{Q}^O_{PM,PM}$$

From the balance equations, we can derive that,

$$\boldsymbol{\pi}_2 = \boldsymbol{\pi}_1\mathbf{H}_{12}; \boldsymbol{\pi}_3 = \boldsymbol{\pi}_1\mathbf{H}_{13}; \boldsymbol{\pi}_4 = \boldsymbol{\pi}_1\mathbf{H}_{14}; \boldsymbol{\pi}_5 = \boldsymbol{\pi}_1\mathbf{H}_{15}; \boldsymbol{\pi}_6 = \boldsymbol{\pi}_1\mathbf{H}_{16}, \tag{8}$$

where

$$\mathbf{H}_{12} = -\mathbf{Q}_{12}\mathbf{Q}_{22}^{-1}; \mathbf{H}_{13} = -\mathbf{Q}_{13}\mathbf{Q}_{33}^{-1}; \mathbf{H}_{14} = -\mathbf{Q}_{14}\mathbf{Q}_{44}^{-1}; \mathbf{H}_{15}$$
$$= -(\mathbf{H}_{12}\mathbf{Q}_{25} + \mathbf{H}_{13}\mathbf{Q}_{35})\mathbf{Q}_{55}^{-1};$$

$$\mathbf{H}_{16} = -(\mathbf{Q}_{16} + \mathbf{H}_{12}\mathbf{Q}_{26})\mathbf{Q}_{66}^{-1}; \boldsymbol{\pi}_1\mathbf{Q}_{11} + \boldsymbol{\pi}_1\mathbf{H}_{12}\mathbf{Q}_{21} + \boldsymbol{\pi}_1\mathbf{H}_{14}\mathbf{Q}_{41} + \boldsymbol{\pi}_1\mathbf{H}_{15}\mathbf{Q}_{51}$$
$$+ \boldsymbol{\pi}_1\mathbf{H}_{16}\mathbf{Q}_{61}$$
$$= \mathbf{0};$$

Finally, from the normalization condition, we obtain $\boldsymbol{\pi}_1$ as,

$$\boldsymbol{\pi}_1 = (1, \mathbf{0})[\mathbf{e} + \mathbf{H}_{12}\mathbf{e} + \mathbf{H}_{13}\mathbf{e} + \mathbf{H}_{14}\mathbf{e} + \mathbf{H}_{15}\mathbf{e} + \mathbf{H}_{16}\mathbf{e}|$$
$$(\mathbf{Q}_{11} + \mathbf{H}_{12}\mathbf{Q}_{21} + \mathbf{H}_{14}\mathbf{Q}_{41} + \mathbf{H}_{15}\mathbf{Q}_{51} + \mathbf{H}_{16}\mathbf{Q}_{61}) *]^{-1}. \tag{9}$$

The stationary distribution for the discrete case is shown in a matrix-algorithmic way in Appendix B3.

## 5. Measures

This section addresses several important metrics in the field of reliability, including availability, reliability, and various mean number of events.

### 5.1. The availability and reliability

The availability is the probability of being operational the system at a determine time $t$, in continuous time, or $\nu$ in discrete time. This measure is obtained from the block-wise transient and stationary distributions, in both the continuous and discrete cases, as given in (6), (7), (8), (9), (14) and (15), respectively. On the other hand, the reliability is the probability of being operational at a certain time before the first failure. The distribution up to first failure is PH distributed. Table 1 shows these measures for the continuous and discrete case in transient and stationary regime.

### 5.2. Mean number of events

The model has been designed by incorporating a structure that facilitates the computation of the average number of events as determined by the MMAP (the proportional number of events per unit of time in a steady state). Table 2 displays this measure for both transient and stationary cases for continuous and discrete distributions. These measures are obtained from the matrix blocks corresponding to the various events of the constructed MMAP, and from the transient and stationary distributions given in (4) and (5) for the respective cases.

## 6. Costs and rewards

The system described is influenced by various events that can generate costs and rewards based on the defined macro-states. On one hand, these costs and rewards arise while the system is either operational or undergoing repair, depending on its temporal evolution. On the other hand, fixed costs are incurred due to events that may occur over time.

### 6.1. Building the cost/reward vector

First, the vector associated with costs and rewards is described and constructed, accounting for periods when the device is operational, broken or in the repair facility, as well as time-dependent costs related to the repairperson vacation policy.

Whenever the system is functioning, a gross profit of $B$ monetary units (m.u.) per unit of time is earned. Conversely, whenever the system is not operational, a cost of $C$ m.u. per unit of time is incurred. Additionally, there is a cost associated with the system being operational, which varies according to the operational level $k=1,\ldots,K$. This cost is represented by the column vector $\mathbf{c}_0^k$ with order $n_k$. Moreover, the vector of expected cost per unit of time due to the external damage while the system is in operation is given by $\mathbf{c}_d$.

If the system is in macro-state $PM$ or $CR$, the repairperson incurs a cost per unit of time depending on the respective repair phases. These costs are represented by the vectors $\mathbf{c}_{PM}$ and $\mathbf{c}_{CR}$, respectively.

The repairperson also generates costs. The cost per unit of time while the repairperson is present at the repair facility (regardless of whether they are actively working) is equal to $H$ m.u. Additionally, if the repairperson is on vacation, the expected cost per unit of time is $F$ m.u.

The average net cost vector for any phase of the system is given by

$$\mathbf{c} = \begin{pmatrix} \mathbf{c}_{O^v} \\ \mathbf{c}_{O^{nv}} \\ \mathbf{c}_{RF} \\ \mathbf{c}_{NRF} \\ \mathbf{c}_{CR} \\ \mathbf{c}_{PM} \end{pmatrix}, \quad (10)$$

where

$$\mathbf{c}_{O^v} = (B - F) \otimes \mathbf{e}_{mtdv} - \mathbf{c}_0 \otimes \mathbf{e}_{tdv} - \mathbf{e}_{mt} \otimes \mathbf{c}_d \otimes \mathbf{e}_v$$

$$\mathbf{c}_{O^{nv}} = (B - F) \otimes \mathbf{e}_{(m-m_K)td} - \mathbf{c}_0^{1,\ldots,K-1} \otimes \mathbf{e}_{td} - \mathbf{e}_{(m-m_K)t} \otimes \mathbf{c}_d,$$

$$\mathbf{c}_{RF} = \mathbf{c}_{NRF} = -(C + F) \otimes \mathbf{e}_{tv},$$

$$\mathbf{c}_{CR} = -(C + H) \otimes \mathbf{e}_{tz_1} - \mathbf{e}_t \otimes \mathbf{cr}_1,$$

$$\mathbf{c}_{PM} = -(C + H) \otimes \mathbf{e}_{tz_2} - \mathbf{e}_t \otimes \mathbf{cr}_2,$$

being

$$\mathbf{c}_0 = \begin{pmatrix} \mathbf{c}_0^1 \\ \vdots \\ \mathbf{c}_0^{K-1} \\ \mathbf{c}_0^K \end{pmatrix},$$

and

$$\mathbf{c}_0^{1,\ldots,K-1} = \begin{pmatrix} \mathbf{c}_0^1 \\ \mathbf{c}_0^2 \\ \vdots \\ \mathbf{c}_0^{K-1} \end{pmatrix}.$$

The mean net profit up to time $t$ in continuous case, or $\nu$ in discrete case, generated by the system, is given by, $\Phi(t) = \int_0^t \mathbf{p}(u)du \cdot \mathbf{c}$ and $\Phi^\nu = \sum_{r=0}^\nu \mathbf{p}^r \cdot \mathbf{c}$ respectively. In stationary regime, this measure can be interpreted as the mean net profit/cost per unit of time $\Phi = \boldsymbol{\pi} \cdot \mathbf{c}$.

### 6.2. Fixed costs according to events and profit measures

Other measures of interest in the evolution of the system, depending on event and not on time in a direct way, are the following,

$G$: fixed cost associated with each return of the repairperson

**Table 1**
Transient distribution and availability and reliability in transient and stationary regime for both continuous and discrete time.

|  | Continuous Case | Discrete Case |
|---|---|---|
| **Availability (transient)** | $A(t) = \mathbf{p}_{E_1}(t) \cdot \mathbf{e} + \mathbf{p}_{E_2}(t) \cdot \mathbf{e}$ | $A^\nu = \mathbf{p}_{E_1}^\nu \cdot \mathbf{e} + \mathbf{p}_{E_2}^\nu \cdot \mathbf{e}$ |
| **Availability (stationary)** | $A = \boldsymbol{\pi}_1 \cdot \mathbf{e} + \boldsymbol{\pi}_2 \cdot \mathbf{e}$ | $A = \boldsymbol{\pi}_1 \cdot \mathbf{e} + \boldsymbol{\pi}_2 \cdot \mathbf{e}$ |
| **Reliability** | $\mathbf{PH}\left((\boldsymbol{\alpha} \otimes \boldsymbol{\pi}_L^c \otimes \boldsymbol{\omega} \otimes \boldsymbol{\upsilon}, \mathbf{0}), \begin{pmatrix} \mathbf{Q}_{11} & \mathbf{Q}_{12} \\ \mathbf{Q}_{21} & \mathbf{Q}_{22} \end{pmatrix}\right)$ | $\mathbf{PH}\left((\boldsymbol{\alpha} \otimes \boldsymbol{\pi}_L^d \otimes \boldsymbol{\omega} \otimes \boldsymbol{\upsilon}, \mathbf{0}), \begin{pmatrix} \mathbf{D}_{11} & \mathbf{D}_{12} \\ \mathbf{D}_{21} & \mathbf{D}_{22} \end{pmatrix}\right)$ |
|  | $R(t) = (\boldsymbol{\alpha} \otimes \boldsymbol{\pi}_L^c \otimes \boldsymbol{\omega} \otimes \boldsymbol{\upsilon}, \mathbf{0})$ | $R^\nu = (\boldsymbol{\alpha} \otimes \boldsymbol{\pi}_L^d \otimes \boldsymbol{\omega} \otimes \boldsymbol{\upsilon}, \mathbf{0}) \cdot \begin{pmatrix} \mathbf{D}_{11} & \mathbf{D}_{12} \\ \mathbf{D}_{21} & \mathbf{D}_{22} \end{pmatrix}^\nu$ |
|  | $\cdot \exp\left(\begin{pmatrix} \mathbf{Q}_{11} & \mathbf{Q}_{12} \\ \mathbf{Q}_{21} & \mathbf{Q}_{22} \end{pmatrix} t\right) \cdot \mathbf{e}$ | $\cdot \left(\mathbf{I} - \begin{pmatrix} \mathbf{D}_{11} & \mathbf{D}_{12} \\ \mathbf{D}_{21} & \mathbf{D}_{22} \end{pmatrix}\right)^{-1} \cdot \begin{pmatrix} \mathbf{D}_{11} & \mathbf{D}_{12} \\ \mathbf{D}_{21} & \mathbf{D}_{22} \end{pmatrix}^0 \cdot \mathbf{e}$ |

**Table 2**
Mean number of events up to a certain time for both continuous and discrete time and both transient and stationary regime.

| Mean number of events up time $\nu$ | Continuous | Discrete |
| --- | --- | --- |
| **Repairable failures** | $\Psi^{RF}(t) = \int_0^t \mathbf{p}(u)du(\mathbf{Q}^{RF} + \mathbf{Q}^{RF+CR}) \cdot \mathbf{e}$ | $\Psi_\nu^{RF} = \sum_{n=0}^\nu \mathbf{p}^n(\mathbf{D}^{RF} + \mathbf{D}^{RF+CR} + \mathbf{D}^{R+RF+CR}) \cdot \mathbf{e}$ |
| (steady-state) | $\Psi^{RF} = \pi(\mathbf{Q}^{RF} + \mathbf{Q}^{RF+CR}) \cdot \mathbf{e}$ | $\Psi^{RF} = \pi(\mathbf{D}^{RF} + \mathbf{D}^{RF+CR} + \mathbf{D}^{R+RF+CR}) \cdot \mathbf{e}$ |
| **Non-repairable failures** | $\Psi^{NRF}(t) = \int_0^t \mathbf{p}(u)du(\mathbf{Q}^{NRF} + \mathbf{Q}^{NRF+NU}) \cdot \mathbf{e}$ | $\Psi_\nu^{RF} = \sum_{n=0}^\nu \mathbf{p}^n(\mathbf{D}^{NRF} + \mathbf{D}^{NRF+NU} + \mathbf{D}^{R+NRF+NU}) \cdot \mathbf{e}$ |
| (steady-state) | $\Psi^{NRF} = \pi(\mathbf{Q}^{NRF} + \mathbf{Q}^{NRF+NU}) \cdot \mathbf{e}$ | $\Psi^{NRF} = \pi(\mathbf{D}^{NRF} + \mathbf{D}^{NRF+NU} + \mathbf{D}^{R+NRF+NU}) \cdot \mathbf{e}$ |
| **Corrective repairs** | $\Psi^{CR}(t) = \int_0^t \mathbf{p}(u)du(\mathbf{Q}^{RF+CR} + \mathbf{Q}^{R+CR}) \cdot \mathbf{e}$ | $\Psi_\nu^{CR} = \sum_{n=0}^\nu \mathbf{p}^n(\mathbf{D}^{RF+CR} + \mathbf{D}^{R+RF+CR} + \mathbf{D}^{R+CR}) \cdot \mathbf{e}$ |
| (steady-state) | $\Psi^{CR} = \pi(\mathbf{Q}^{RF+CR} + \mathbf{Q}^{R+CR}) \cdot \mathbf{e}$ | $\Psi^{CR} = \pi(\mathbf{D}^{RF+CR} + \mathbf{D}^{R+RF+CR} + \mathbf{D}^{R+CR}) \cdot \mathbf{e}$ |
| **Preventive maintenance** | $\Psi^{PM}(t) = \int_0^t \mathbf{p}(u)du(\mathbf{Q}^{PM} + \mathbf{Q}^{R+PM}) \cdot \mathbf{e}$ | $\Psi_\nu^{PM} = \sum_{n=0}^\nu \mathbf{p}^n(\mathbf{D}^{PM} + \mathbf{D}^{R+PM}) \cdot \mathbf{e}$ |
| (steady-state) | $\Psi^{PM} = \pi(\mathbf{Q}^{PM} + \mathbf{Q}^{R+PM}) \cdot \mathbf{e}$ | $\Psi^{PM} = \pi(\mathbf{D}^{PM} + \mathbf{D}^{R+PM}) \cdot \mathbf{e}$ |
| **Returns** | $\Psi^{R}(t) = \int_0^t \mathbf{p}(u)du \begin{pmatrix} \mathbf{Q}^{R} + \mathbf{Q}^{R+CR} + \mathbf{Q}^{R+PM} \\ +\mathbf{Q}^{R+NU} + \mathbf{Q}^{R+NVP} \end{pmatrix} \cdot \mathbf{e}$ | $\Psi_\nu^{R} = \sum_{n=0}^\nu \mathbf{p}^n \begin{pmatrix} \mathbf{D}^{R} + \mathbf{D}^{R+CR} + \mathbf{D}^{R+NVP} + \\ \mathbf{D}^{R+NU} + \mathbf{D}^{R+RF+CR} + \mathbf{D}^{R+NRF+NU} \end{pmatrix} \cdot \mathbf{e}$ |
| (steady-state) | $\Psi^{R} = \pi \begin{pmatrix} \mathbf{Q}^{R} + \mathbf{Q}^{R+CR} + \mathbf{Q}^{R+PM} \\ +\mathbf{Q}^{R+NU} + \mathbf{Q}^{R+NVP} \end{pmatrix} \cdot \mathbf{e}$ | $\Psi_\nu^{R} = \pi \begin{pmatrix} \mathbf{D}^{R} + \mathbf{D}^{R+CR} + \mathbf{D}^{R+NVP} + \mathbf{D}^{R+NU} \\ +\mathbf{D}^{R+RF+CR} + \mathbf{D}^{R+NRF+NU} \end{pmatrix} \cdot \mathbf{e}$ |
| **New units** | $\Psi^{NU}(t) = \int_0^t \mathbf{p}(u)du(\mathbf{Q}^{NRF+NU} + \mathbf{Q}^{R+NU}) \cdot \mathbf{e}$ | $\Psi_\nu^{NU} = \sum_{n=0}^\nu \mathbf{p}^n(\mathbf{D}^{NRF+NU} + \mathbf{D}^{R+NRF+NU} + \mathbf{D}^{R+NU}) \cdot \mathbf{e}$ |
| (steady-state) | $\Psi^{NU} = \pi(\mathbf{Q}^{NRF+NU} + \mathbf{Q}^{R+NU}) \cdot \mathbf{e}$ | $\Psi^{NU} = \pi(\mathbf{D}^{NRF+NU} + \mathbf{D}^{R+NRF+NU} + \mathbf{D}^{R+NU}) \cdot \mathbf{e}$ |
| **New vacation periods** | $\Psi^{NVP}(t) = \int_0^t \mathbf{p}(u)du \mathbf{Q}^{R+NVP} \cdot \mathbf{e}$ | $\Psi_\nu^{NVP} = \sum_{n=0}^\nu \mathbf{p}^n \mathbf{D}^{R+NVP} \cdot \mathbf{e}$ |
| (steady-state) | $\Psi^{NVP} = \pi \mathbf{Q}^{R+NVP} \cdot \mathbf{e}$ | $\Psi^{NVP} = \pi \mathbf{D}^{R+NVP} \cdot \mathbf{e}$ |

(regardless of whether they stay or not).

*fcr*: fixed cost each time that the unit begins a corrective repair.

*fpm*: fixed cost each time that the unit begins a preventive maintenance.

*fnu*: cost for a new unit.

To evaluate the total average cost up to a certain time, the following measure, mean net total profit up to a certain time, is defined in continuous, and discrete time.

*Continuous Case:*

$$\Lambda(t) = \Phi(t) - (1 + \Psi^{NU}(t)) \cdot fnu - \Psi^{CR}(t) \cdot fcr \\ -\Psi^{PM}(t) \cdot fpm - \Psi^{R}(t) \cdot G \quad (11)$$

*Discrete Case*

$$\Lambda^\nu = \Phi^\nu - (1 + \Psi_\nu^{NU}) \cdot fnu - \Psi_\nu^{CR} \cdot fcr \\ -\Psi_n^{PM} \cdot fpm - \Psi_\nu^{R} \cdot G$$

This measure in stationary regime is given below. It can be interpreted as the net total profit per unit of time.

*Continuous Case*

$$\Lambda = \lim_{t\to\infty}\frac{\Lambda(t)}{t} = \Phi - \Psi^{NU} \cdot fnu - \Psi^{CR} \cdot fcr \\ -\Psi^{PM} \cdot fpm - \Psi^{R} \cdot G \quad (12)$$

*Discrete Case*

$$\Lambda = \lim_{\nu\to\infty}\frac{\Lambda^\nu}{\nu+1} = \Phi - \Psi^{NU} \cdot fnu - \Psi^{CR} \cdot fcr \\ -\Psi^{PM} \cdot fpm - \Psi^{R} \cdot G$$

## 7. Optimizing

A fundamental aspect of analyzing system behaviour is the optimization of the Bernoulli vacation policy. Ideally, optimization aims for a system with maximum availability and maximum profit; however, the higher the availability, the lower the resulting profit, and vice versa. In this context, multiple optimization possibilities are considered:

1. Maximizing net profit per unit of time.
2. Maximizing availability.
3. Performing a multi-objective Pareto analysis, considering both net profit per unit time and availability in steady-state conditions. Once the Pareto front is obtained, a possible selection criterion is to choose the point closest to an ideal point.

These options will be considered in the numerical example.

In Section 6.2, the net profit in the stationary regime has been calculated in equation (12). Meanwhile, Section 5.1 presents the availability in the stationary regime (Table 1). Both metrics have been analyzed for the discrete and continuous cases. These two measures depend on the vacation policy, the vacation duration, $(\nu, \mathbf{V})$, and the probabilities of starting a new vacation period upon return and observing a degradation level $k$, $p_k$, for $k = 1, \ldots, K-1$.

Point 3, the multi-objective Pareto analysis, is carried out as follows. If the objective functions to be maximized are $f_1$, the net profit per unit of time under steady-state conditions, and $f_2$, the availability in steady-state conditions, then the ideal point is a vector $z$, optimized over the feasible values in the Pareto analysis,

$$z_i^* = \max_{\substack{(\nu, \mathbf{V}) \\ p_k; k=1,\ldots,K-1}} f_i.$$

The solution closest to the ideal point, which will be selected from the Pareto front, is the one obtained by applying the Euclidean norm after normalization,

$$\min_{\substack{(\nu, \mathbf{V}) \\ p_k; k=1,\ldots,K-1}} \| f_{norm} - z^* \|_2 \quad (13)$$

## 8. Numerical example

The numerical example presented in this section could be motivated by the continuous operation of a high-precision CNC (Computer Numerical Control) milling machine used in the aerospace manufacturing industry. These machines are composed of complex subsystems and operate continuously in demanding production environments, making them susceptible to gradual internal wear, sudden failures, and external disruptions. The system evolves over time with three degradation levels: normal operation, moderate degradation, and severe degradation. Additionally, the device is also exposed to external shocks, such as, voltage fluctuations, power surges, unexpected material inconsistencies, and abrupt command interruptions from the control system.

A maintenance technician is responsible for a fleet of such machines but is not always available due to multitasking responsibilities or scheduled leave. When the repairperson is present and a repairable failure is detected, the machine enters corrective repair. If the technician observes that the system is operating at level 3, which represents a critical level, preventive maintenance is carried out to avoid catastrophic failure. Both corrective and preventive repair follow phase-type distributions.

The primary objective is to optimize the vacation time of the repairperson and the reassignment policy, ensuring high machine availability while minimizing operational and economic costs. This is particularly relevant in industries where downtime is extremely expensive and predictive maintenance strategies are essential.

### 8.1. Phase-type embedded times

A unitary system is considered, where internal wear is partitioned into three performance or degradation levels, each with 2, 3, and 2 internal operational phases, respectively. The last one represents the critical level. The phase-type distributions embedded in the system, as well as the different probabilistic structures, are shown in Table 3.

The transition intensities between phases are regulated by matrix $\mathbf{T}$. From each of these phases, a failure may occur, either repairable or non-repairable, as indicated by the vectors $\mathbf{T}_r^0$ and $\mathbf{T}_{nr}^0$, respectively. Thus, it can be observed that from degradation level 1, no failure can occur directly; from level 2, only repairable failures can occur; and from level 3, critical level, only non-repairable failures can occur.

The average residence time in each level, without considering external factors, is 100, 11.375, and 1.4 units of time (u.t.), respectively.

The behaviour of the unit is also influenced by external shocks governed by a phase-type process with inter-event times that can produce shocks represented by $(\boldsymbol{\gamma}, \mathbf{L})$. The mean time between two consecutive events is 5 u.m. This shock may result in an internal modification matrix ($\mathbf{W}$), a repairable internal failure ($\mathbf{W}_r^0$), or a non-repairable internal failure ($\mathbf{W}_{nr}^0$). In cases where no failure occurs, the third external shock will result in a non-repairable failure in the unit ($\mathbf{C}$). A high-intensity external shock causes a non-repairable failure in the unit, and we assume these shocks occur with probability $\omega^0 = 0.2$.

When the repairperson observes a repairable failure, the unit enters the repair channel to undergo corrective repair. The time for this repair follows a phase-type distribution represented by $(\boldsymbol{\beta}^1, \mathbf{S}_1)$, with a mean corrective repair time of 6.4384 u.t. On the other hand, to prevent major system issues and their corresponding consequences, when the technician observes that the unit is at the third internal functioning level, the unit enters the repair channel to undergo preventive maintenance. The time for preventive maintenance follows a phase-type distribution represented by $(\boldsymbol{\beta}^2, \mathbf{S}_2)$, with a mean time of 1.1645 u.t.

### 8.2. Benefits and costs

As described in Section 6, the system incurs benefits and costs throughout its operation. It is assumed that while the system is operational, it generates a benefit of $B = 15$ monetary units (m.u.) per unit of time. Consequently, the loss per unit of time when the system is not operational is considered equal, $C=15$ m.u. Additionally, depending on the functioning level of the unit, there is a cost (or reduction in benefit) as follows: If the system is operating at level 1, there are no losses. If it operates at level 2, there is a loss of 2 u.m. per unit of time, and if it operates at level 3, the loss increases to 10 u.m. Therefore, the associated vector is $\mathbf{c}_0 = (0, 0, 2, 2, 2, 10, 10)'$.

**Table 3**
Matrices embedded in the system.

$\boldsymbol{\alpha} = (1, 0, 0, 0, 0, 0, 0)$

$$\mathbf{T} = \begin{pmatrix} -0.2 & 0.2 & 0 & 0 & 0 & 0 & 0 \\ 0.18 & -0.2 & 0.02 & 0 & 0 & 0 & 0 \\ 0 & 0 & -0.5 & 0.2 & 0.3 & 0 & 0 \\ 0 & 0 & 0.25 & -0.45 & 0.15 & 0 & 0 \\ 0 & 0 & 0 & 0.2 & -0.4 & 0.02 & 0 \\ 0 & 0 & 0 & 0 & 0 & -1 & 0.8 \\ 0 & 0 & 0 & 0 & 0 & 0 & -2 \end{pmatrix}$$

$$\mathbf{T}_r^0 = \begin{pmatrix} 0 \\ 0 \\ 0 \\ 0.05 \\ 0.18 \\ 0 \\ 0 \end{pmatrix} ; \mathbf{T}_{nr}^0 = \begin{pmatrix} 0 \\ 0 \\ 0 \\ 0 \\ 0 \\ 0.2 \\ 2 \end{pmatrix}$$

$\boldsymbol{\gamma} = (1, 0)$

$$\mathbf{L} = \begin{pmatrix} -0.8 & 0.8 \\ 0.1 & -0.4 \end{pmatrix}$$

$\boldsymbol{\omega} = (1, 0, 0)$

$$\mathbf{C} = \begin{pmatrix} 0 & 1 & 0 \\ 0 & 0 & 1 \\ 0 & 0 & 0 \end{pmatrix}$$

$$\mathbf{W} = \begin{pmatrix} 0.3 & 0.7 & 0 & 0 & 0 & 0 & 0 \\ 0.1 & 0.3 & 0.6 & 0 & 0 & 0 & 0 \\ 0 & 0 & 0.4 & 0.5 & 0.1 & 0 & 0 \\ 0 & 0 & 0 & 0.3 & 0.6 & 0 & 0 \\ 0 & 0 & 0 & 0 & 0.3 & 0.5 & 0 \\ 0 & 0 & 0 & 0 & 0 & 0 & 0.7 \\ 0 & 0 & 0 & 0 & 0 & 0 & 0.5 \end{pmatrix}$$

$$\mathbf{W}_r^0 = \begin{pmatrix} 0 \\ 0 \\ 0 \\ 0.1 \\ 0.2 \\ 0 \\ 0 \end{pmatrix} ; \mathbf{W}_{nr}^0 = \begin{pmatrix} 0 \\ 0 \\ 0 \\ 0 \\ 0 \\ 0.3 \\ 0.5 \end{pmatrix}$$

$\boldsymbol{\beta}^1 = (1, 0, 0)$

$$\mathbf{S}_1 = \begin{pmatrix} -0.9 & 0.5 & 0.3 \\ 0.2 & -0.6 & 0.1 \\ 0 & 0.1 & -0.2 \end{pmatrix}$$

$\boldsymbol{\beta}^2 = (1, 0, 0)$

$$\mathbf{S}_2 = \begin{pmatrix} -0.9 & 0.02 & 0.01 \\ 0.1 & -0.8 & 0.05 \\ 0 & 0.05 & -0.6 \end{pmatrix}$$

On the other hand, the system is considered to endure up to two external shocks before being replaced. In this case, these external shocks do not incur any costs, so $\mathbf{c}_d = (0, 0, 0)'$.

When events occur, fixed costs are incurred, with the failure-repair policy defined as follows: The fixed cost each time the unit enters the repair channel for corrective repair is $fcr = 20$, while for preventive maintenance tasks, it is $fpm = 2$.

On the other hand, corrective repair and preventive maintenance tasks are carried out through three phases in each case. The average cost per unit of time in each of these phases is given by $\mathbf{cr}_1 = (10, 20, 30)'$ and $\mathbf{cr}_2 = (1, 2, 3)'$ for corrective repair and preventive maintenance, respectively.

The repair technician also incurs fixed costs whether working, in the repair channel, or on vacation. Specifically, the cost per unit of time that the technician is in the repair channel is $H = 3.5$, and the cost per unit of time when the technician is on vacation is $F = 1$. Additionally, the technician's returns to work also incur a cost of $G = 0.5$ u.m. per return.

Finally, the cost of each new unit is $fnu = 100$ m.u.

All these values have been considered for the construction of the expected net benefit/cost vector per unit of time, according to the operational phase and degradation level, as given in equation (12).

*8.3. Optimizing the system*

With the system's repair and operation times, as well as the incorporated benefits and costs, the objective is to optimize the vacation policy for the repair channel. For the optimization process, the vacation time is modeled as a Coxian distribution of order 3 with a general representation $(\boldsymbol{v}, \mathbf{V})$. Thus,

$$\boldsymbol{v} = (1, 0, 0); \mathbf{V} = \begin{pmatrix} -V_1 & V_2 & 0 \\ 0 & -V_3 & V_4 \\ 0 & 0 & -V_5 \end{pmatrix}$$

Furthermore, the aim is to optimize the probability $1\text{-}p_1$ that the repairperson remains in the system upon returning from vacation and observing a level 1 operational state, and $1\text{-}p_2$ for a level 2 operational state.

In total, seven parameters are optimized. As mentioned in Section 7, a multi-objective Pareto analysis is performed, optimizing both the net benefit and the steady-state availability.

Upon completing the analysis, the Pareto front shown in Fig. 2 is obtained.

From the set of Pareto-optimal solutions, several selection criteria may be considered:

- First, one may select the solution with the highest net profit (Model 1: red star in Fig. 2).

Note that this optimization takes into account both costs and the system performance, which is influenced by the optimal vacation time.

- Second, the solution with the highest steady-state availability can be selected (Model 3: blue star in Fig. 2).

This criterion does not consider costs in the optimization, only the system's operational performance, which will later be reflected in the analysis.

- Third, Model 2 (black star in Fig. 2) corresponds to the criterion described in equation (13). This approach considered an ideal point that simultaneously maximizes both objectives, (0.2734, 0.9187), and selects, from the feasible solutions, the one with the shortest normalized distance to this point. Under this criterion, the optimal solution yields a net profit per unit of time equal to 0.0164 and a steady-state availability equal to 0.9168.

A comparative analysis using the three criteria has been carried out in the development of the numerical example. The optimal values for the three models are presented in Table 4.

To interpret the procedure based on these results, we focus on Model 2, although, as previously mentioned, a comparative analysis of the three models will be conducted later.

The phase-type representation of the vacation time is expressed. From this distribution, the repairperson's return times can be simulated. Upon the repairperson's return, the following situations may arise, with actions taken as described:

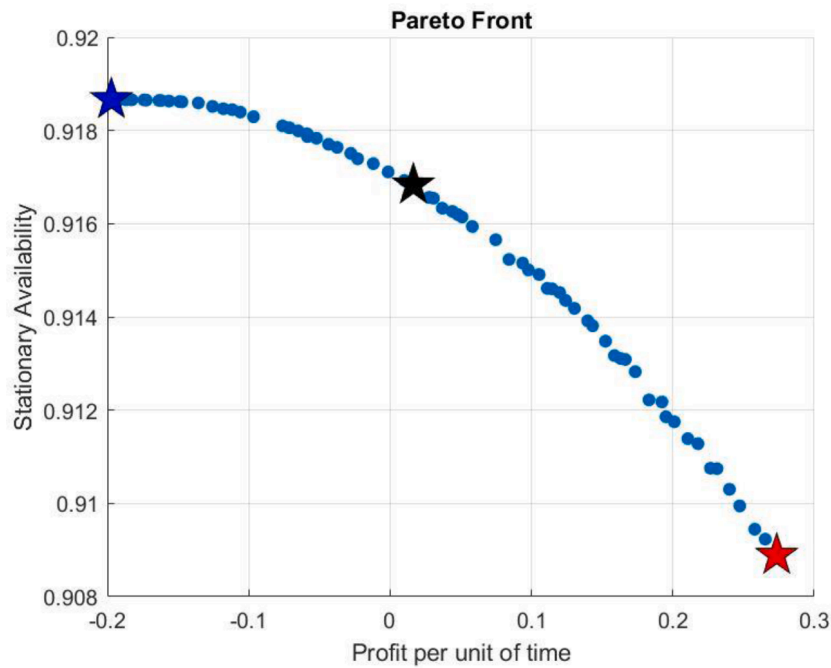

**Fig. 2.** Pareto front. Profit per unit of time vs. steady-state availability.

**Table 4**
Selected Optimal Values from the Pareto Front.

| | Vacation distribution | Prob. $p_1$ | Prob. $p_2$ | Opt. profit | Opt. Availability |
|---|---|---|---|---|---|
| Model 1 | $\mathbf{V} = \begin{pmatrix} -10.1881 & 10.1659 & 0 \\ 0 & -10.1855 & 9.8288 \\ 0 & 0 & -8.3987 \end{pmatrix}$ | 0.9999 | 0.5089 | 0.2734 | 0.9089 |
| Model 2 | $\mathbf{V} = \begin{pmatrix} -10.2026 & 10.1463 & 0 \\ 0 & -10.1936 & 9.8266 \\ 0 & 0 & -8.4319 \end{pmatrix}$ | 0.9153 | 0.5088 | 0.0164 | 0.9168 |
| Model 3 | $\mathbf{V} = \begin{pmatrix} -959.2034 & 2.1422 & 0 \\ 0 & -634.2397 & 178.3713 \\ 0 & 0 & -390.6219 \end{pmatrix}$ | 0.0379 | 0.3374 | -0.1972 | 0.9187 |

1. The repairperson observes the first two operational phases, level 1. In this case, with probability 0.9153, a new vacation period begins.
2. The repairperson observes operational phases 3, 4, or 5, level 2, in which case they decide to remain idle in the repair channel with probability 0.4912.
3. The repairperson observes the last two operational phases, critical level 3. In this case, they begin performing preventive maintenance.
4. The repairperson observes that the unit is in a repairable failure state. Corrective repair is initiated.
5. The repairperson observes that the unit is in a non-repairable failure state, and the unit is replaced. In this case, a new vacation period begins.

### 8.4. System performance measures post-optimization

Once the system has been optimized, several relevant performance measures have been computed for all models. The proportion of time the system spends in each macro-state under steady-state conditions has been determined based on equations (8) and (9).

As expected, the highest proportion of operational time is achieved with Model 3, which explicitly optimizes this measure, reaching a value of 0.9187, primarily due to the repairperson remaining in the repair facility, since costs are not considered in this model.

Conversely, the lowest operational time corresponds to Model 1, which optimizes only the net profit, with a value of 0.9089. For Model 2, which balances both objectives, the operational proportion is 0.9168.

All three models exhibit similar proportions of time during which the repairperson is operational. However, the proportion of time the repairperson remains idle is highest in Model 3, with a value of 0.9186. Table 5 presents the steady-state time proportions for each macro-state across the different models.

*Availability*

The evolution of the optimized system's availability has also been analyzed. Fig. 3 shows the system's availability converging to a stationary values of 0.9089, 0.9168 and 0.9187 for Model 1, Model 2 and Model 3, respectively.

*Reliability*

The system's transient reliability has also been evaluated. This reliability is defined as the probability that the unit is operational at a given time before failing for the first time due to any cause, as described in Table 1. The operational time for each model follows a phase-type distribution, with mean values of 13.3705, 36.8556 and 61.0399 respectively. Fig. 4 shows the reliability function for the different models.

*Proportion of Events*

Thanks to the developed structure of the model using the MMAP approach, it is possible to compute the average number of events up to a given time as detailed in Table 2 of Section 5.2. Thus, it is straightforward to compute the average proportion of events per unit of time up to a certain time, $\Psi^Y(t)/t$, for event $Y$. Fig. 5, 6 and 7 illustrate the evolution of these event proportions over time, along with their corresponding steady-state values ($\Psi^Y$) for each model.

The main difference arises in the cases where the repairperson returns to remain in the repair channel, and when returns occur to begin a new vacation period.

### 8.5. Profitability behaviour

An important aspect to analyze is the economic performance of the optimized system. The first metric considered is the average profit per unit of time up to a certain time, denoted as $\Lambda(t)/t$, as well as the expected cumulative net profit up to time $t$, $\Lambda(t)$, as defined in equation (11).

From Fig. 8, we observe a rapid recovery from initial losses, followed by stabilization.

- For Model 1 (which optimizes net profit per unit of time), the system begins to generate profit at time 155.7316, eventually stabilizing at a steady-state value of 0.2734.
- The balanced Model 2 also shows a growing profit trajectory but begins to generate positive returns later, starting at time 2646.569, and stabilizes at a lower value of 0.0164.
- Model 3, which optimizes availability without considering costs or profits, never generates profit. Its expected steady-state value is –0.1972, indicating a consistent loss per unit of time.

Fig. 9 displays the cumulative profit. A rapid increase can be observed for Model 1, a more moderate growth for Model 2, and a decline starting at approximately time 20 for Model 3, indicating increasing losses in the latter case.

### 9. Conclusions

In this work, a complex single-unit system is modeled using an novel algorithmic-matrix approach, enabling straightforward computational implementation and interpretation of results. The system is subject to multiple events, including internal failures (repairable or not), external shocks with various consequences, and preventive maintenance. The internal performance of the unit is partitioned into an indeterminate number of levels, each affecting its performance. The system modeling is based on a new Markovian Arrival Process with Marked arrivals. This structure facilitates efficient algorithmization of the model and the derivation of numerous performance metrics related to the defined events.

A novel Bernoulli-type vacation policy is implemented in the repair facility. In this way, the repairperson may be present or absent, allowing for the execution of various tasks. For enhanced efficiency and profitability, if the repairperson observes a high level of degradation, preventive maintenance tasks are performed. Conversely, if no significant degradation is observed, the repairer remains in the repair channel with a certain probability, depending on the operational level. Costs and benefits are incorporated into the system, associated with operational levels, damages, repairs, and preventive maintenance. The system is optimized by determining the optimal vacation policy based on benefits and operational performance.

**Table 5**
Proportional time in each macro-state.

| | $O^v$ | $O^{nv}$ | RF | NRF | CR | PM |
|---|---|---|---|---|---|---|
| Model 1 | 0.7678 | 0.1410 | 0.0001 | 0.0106 | 0.0771 | 0.0034 |
| Model 2 | 0.2474 | 0.6694 | 0.0000 | 0.0020 | 0.0777 | 0.0034 |
| Model 3 | 0.0001 | 0.9186 | 0.0000 | 0.0000 | 0.0779 | 0.0034 |

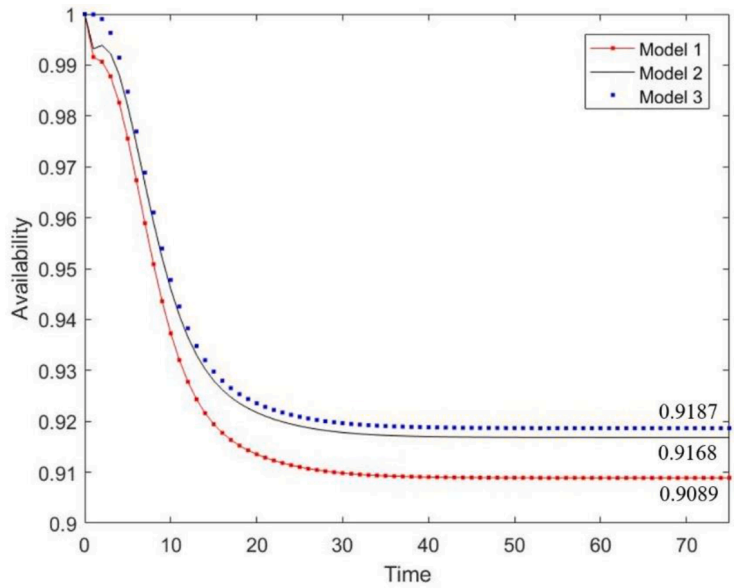

**Fig. 3.** Availability of the system.

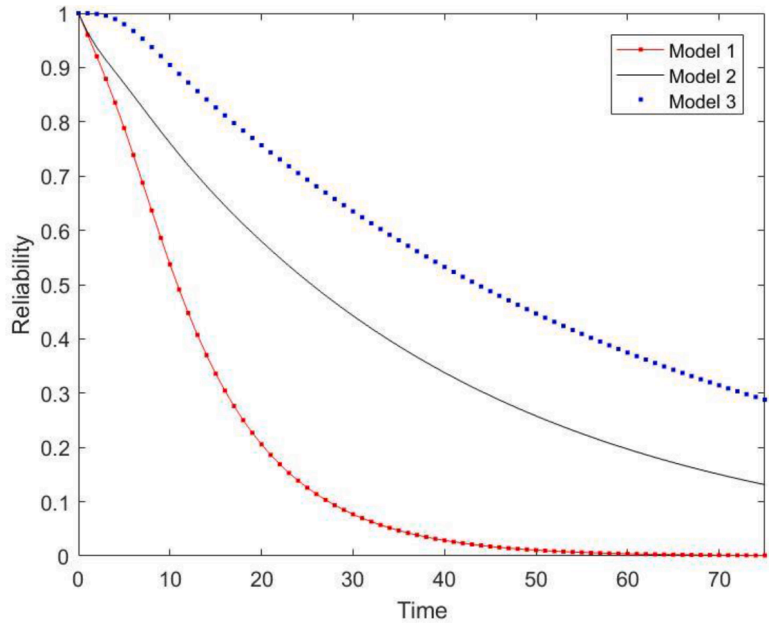

**Fig. 4.** Reliability of the system.

All the aforementioned elements are illustrated through a numerical example in which the system is optimized, performance metrics are obtained, and conclusions are drawn, demonstrating the versatility of the model. For the optimization, a methodology based on multi-objective Pareto analysis is developed, applying multiple criteria; optimization of the expected net benefit-cost per unit of time in the steady-state regime, the closest-point-to-ideal criterion for optimal selection, and selection of optimal values that maximize steady-state availability. A comparative analysis of the three models is conducted, examining both system performance and economic behaviour in each case.

The choice of the optimal model from the Pareto front is crucial for the subsequent analysis of the performance-benefit relationship of the system. The selection criterion of choosing the point closest to the ideal point (maximum availability and maximum profit) appears to be the best option. When comparing the model that maximizes steady-state availability with the balanced system, it is observed that the increase in steady-state availability is only 0.21%, with no economic benefit per unit of time in return. On the other hand, if the model that maximizes only benefits is considered, the steady-state benefit per unit of time increases significantly compared to the balanced model, while availability decreases by 0.86%. Other measures presented in this work may be considered in a comprehensive analysis to support decision-making in each specific case.

The methodology developed in this work is implemented using matrix-based structures, allowing both the exact probabilistic model and the associated performance measures to be formally obtained in an algorithmic and matrix-oriented manner, regardless of the number of operational levels. While this is a significant advantage, it is also true that as the number of levels increases, the computational cost will likewise increase in practice. To mitigate this issue, various techniques can be considered: grouping or eliminating low-probability states, limiting the maximum number of levels, using known phase-type distributions to fix transitions, or applying stochastic simulation techniques.

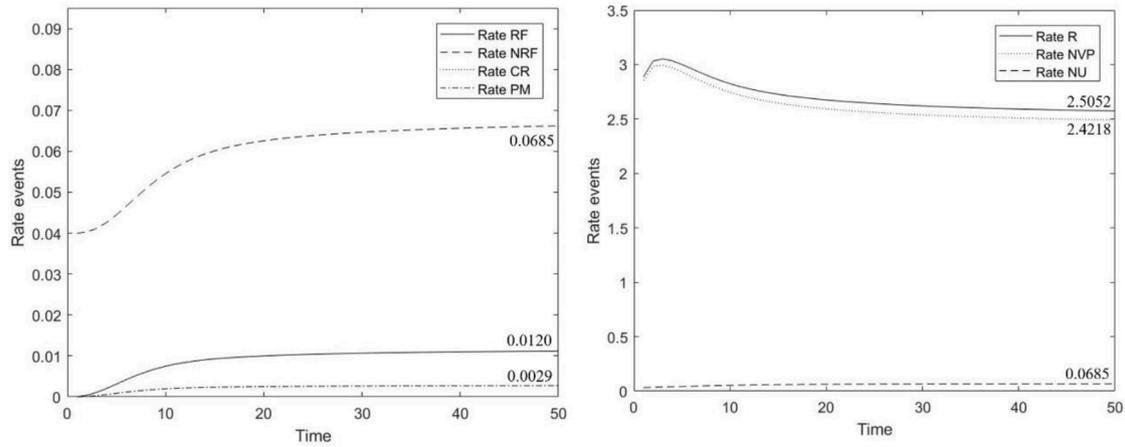

**Fig. 5.** Proportion of events over time and the stationary value for Model 1.

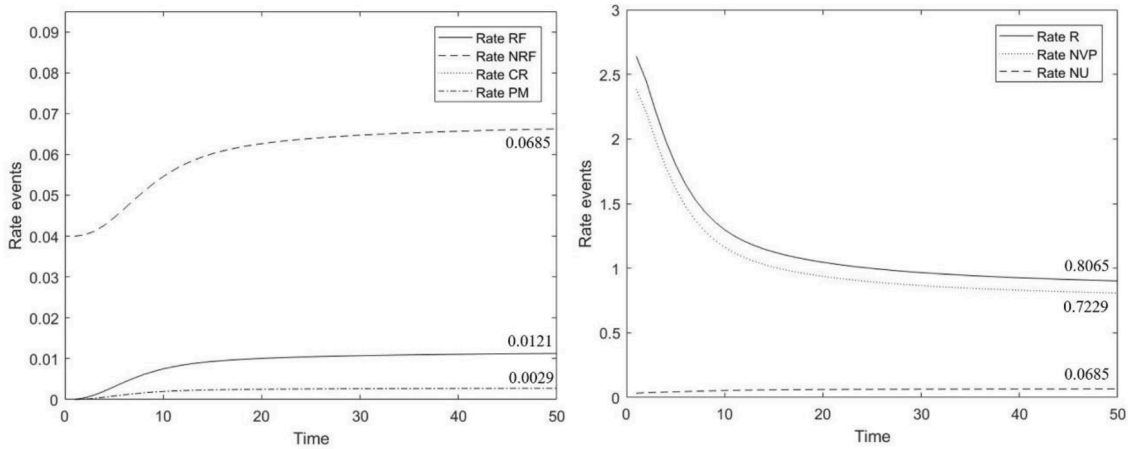

**Fig. 6.** Proportion of events over time and the stationary value for Model 2.

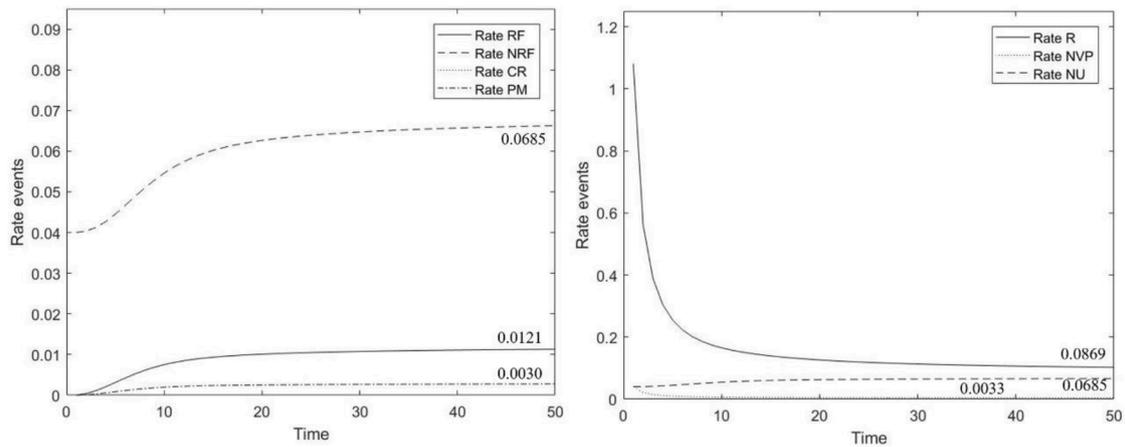

**Fig. 7.** Proportion of events over time and the stationary value for Model 3.

The modeling, as well as the multiple measures, have been conducted for systems evolving in both continuous and discrete time, in both transient and steady-state regimes. All computations and results presented in this work were implemented in MATLAB, yielding accurate outcomes and optimal performance. Evolutionary algorithms based on genetic methodologies were used for optimal solution selection.

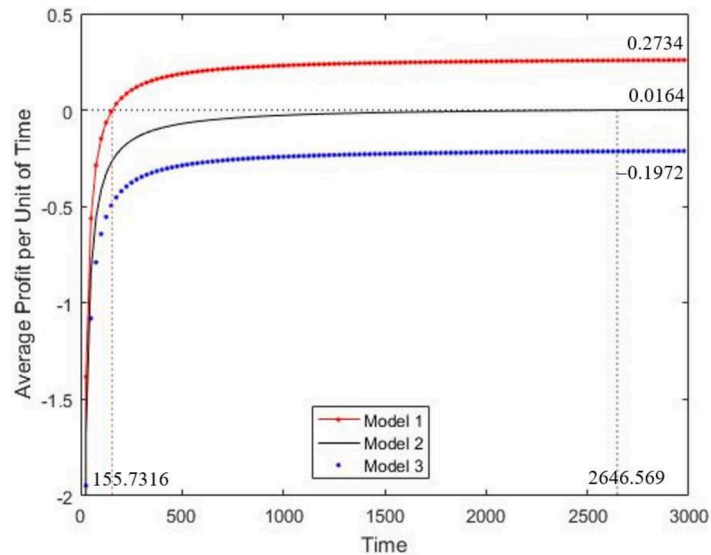

**Fig. 8.** Average net profit per unit of time over time.

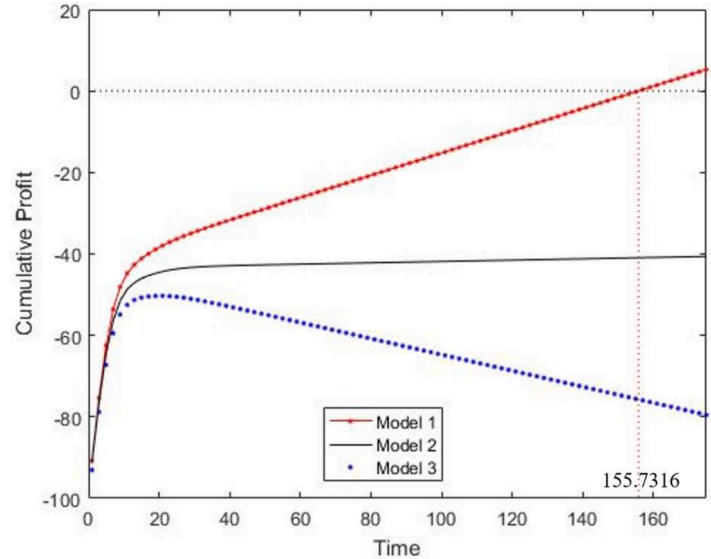

**Fig. 9.** Cumulative net profit over time.

### Declaration of generative AI and AI-assisted technologies in the writing process

During the preparation of this work the authors used webpage chatgpt.com in order to improve language and readability, with caution. After using this webpage, the authors reviewed and edited the content as needed and take full responsibility for the content of the publication.

**Juan Eloy Ruiz Castro:** Conceptualization, Methodology, Software, Formal analysis, Investigation, Writing - Original Draft, Writing - Review & Editing, Supervision, Funding acquisition.

**Hugo Alaín Zapata-Ceballos:** Methodology, Software, Formal analysis, Investigation, Writing - Original Draft, Writing - Review & Editing.

### CRediT authorship contribution statement

**Juan Eloy Ruiz-Castro:** Writing – review & editing, Writing – original draft, Supervision, Software, Resources, Project administration, Methodology, Investigation, Funding acquisition, Formal analysis, Conceptualization. **Hugo Alaín Zapata-Ceballos:** Writing – review & editing, Writing – original draft, Software, Methodology, Investigation, Formal analysis.

### Declaration of competing interest

The authors declare the following financial interests/personal relationships which may be considered as potential competing interests:

Juan Eloy Ruiz-Castro reports financial support was provided by Government of Andalusia. Juan Eloy Ruiz-Castro reports financial support was provided by Spain Ministry of Science and Innovation. Juan Eloy Ruiz-Castro reports financial support was provided by IMAG-María de Maeztu. Juan Eloy Ruiz-Castro reports a relationship with Government of Andalusia that includes: funding grants. Juan Eloy Ruiz-Castro reports a relationship with Spain Ministry of Science and Innovation that includes: funding grants. Juan Eloy Ruiz-Castro reports a relationship with IMAG-María de Maeztu that includes: funding grants. If there are other authors, they declare that they have no known competing financial interests or personal relationships that could have appeared to influence

## Acknowledgemets

This paper is partially supported by the project FQM-307 of the Government of Andalusia (Spain) and by the project PID2023-149087NB-I00 of the Spanish Ministry of Science and Innovation (also supported by the FEDER programme). Additionally, the first author would like to express their gratitude for financial support by the IMAG–María de Maeztu grant CEX2020-001105-M/AEI/10.13039/501100011033.

## APPENDIX A

This appendix contains the rest of the block-matrices for the system in the continuous case, as introduced in Section 3.3.5. For its development, the matrix functions defined in (1), (2), and (3) have been used.

Matrices $\mathbf{Q}^{RF+CR}$ and $\mathbf{Q}^{NRF+NU}$

The matrices $\mathbf{Q}^{RF+CR}$ and $\mathbf{Q}^{NRF+NU}$ govern the transition when the repairperson is not on vacation, the unit is working, and it undergoes a repairable or non-repairable failure, respectively. It begins a corrective repair because the repairperson was in the workplace or the unit is replaced, respectively.

$$\mathbf{Q}^{RF+CR} = \begin{array}{c} O^v \\ O^{nv} \\ RF \\ NRF \\ CR \\ PM \end{array} \begin{pmatrix} 0 & 0 & 0 & 0 & 0 & 0 \\ 0 & 0 & 0 & 0 & \mathbf{Q}^{RF+CR}_{O^{nv},CR} & 0 \\ 0 & 0 & 0 & 0 & 0 & 0 \\ 0 & 0 & 0 & 0 & 0 & 0 \\ 0 & 0 & 0 & 0 & 0 & 0 \\ 0 & 0 & 0 & 0 & 0 & 0 \end{pmatrix},$$

where

$$\mathbf{Q}^{RF+CR}_{O^{nv},CR} = \mathbf{H}_{RF}\left(\mathbf{U}_{1,\ldots,K-1}\right) \otimes \boldsymbol{\beta}^1.$$

The repairperson is idle at the workplace, and a repairable failure occurs. Before the failure, the system is at a level below critical (otherwise, it would undergo preventive maintenance given that the repairperson is in the workplace) and proceeds to corrective repair.

Matrix $\mathbf{Q}^{NRF+NU}$

$$\mathbf{Q}^{NRF+NU} = \begin{array}{c} O^v \\ O^{nv} \\ RF \\ NRF \\ CR \\ PM \end{array} \begin{pmatrix} 0 & 0 & 0 & 0 & 0 & 0 \\ \mathbf{Q}^{NRF+NU}_{O^{nv},O^v} & 0 & 0 & 0 & 0 & 0 \\ 0 & 0 & 0 & 0 & 0 & 0 \\ 0 & 0 & 0 & 0 & 0 & 0 \\ 0 & 0 & 0 & 0 & 0 & 0 \\ 0 & 0 & 0 & 0 & 0 & 0 \end{pmatrix},$$

where

$$\mathbf{Q}^{NRF+NU}_{O^{nv},O^v} = \mathbf{H}_{NRF}\left(\mathbf{U}_{1,\ldots,K-1}, \boldsymbol{\alpha}, \boldsymbol{\omega}\right) \otimes \boldsymbol{\upsilon}.$$

The repairer is at the workplace, a non-repairable failure occurs from a level below critical, and the unit is replaced with a new one with an initial internal behaviour distribution ($\boldsymbol{\alpha}$) and initial external damage ($\boldsymbol{\omega}$). The repairperson begins a vacation period with an initial distribution ($\boldsymbol{\upsilon}$).

Matrix $\mathbf{Q}^{RF}$

This matrix contains the transition intensities to the macro-state *RF*, corresponding to a repairable failure. This is only possible when a repairable failure occurs, and the repairperson is not at their workplace (on vacation),

$$\mathbf{Q}^{RF} = \begin{array}{c} O^v \\ O^{nv} \\ RF \\ NRF \\ CR \\ PM \end{array} \begin{pmatrix} 0 & 0 & \mathbf{Q}^{RF}_{O^v,RF} & 0 & 0 & 0 \\ 0 & 0 & 0 & 0 & 0 & 0 \\ 0 & 0 & 0 & 0 & 0 & 0 \\ 0 & 0 & 0 & 0 & 0 & 0 \\ 0 & 0 & 0 & 0 & 0 & 0 \\ 0 & 0 & 0 & 0 & 0 & 0 \end{pmatrix},$$

where

$$\mathbf{Q}^{RF}_{O^v,RF} = \mathbf{H}_{RF}\left(\mathbf{U}_{1,\ldots,K}\right) \otimes \mathbf{I} = \mathbf{H}_{RF}(\mathbf{I}) \otimes \mathbf{I}.$$

A repairable failure occurs regardless of the internal level of the system ($\mathbf{H}_{RF}(\mathbf{I})$), and there is no modification in transitions during vacation time ($\mathbf{I}$).

Matrix $\mathbf{Q}^{NRF}$

This matrix contains the transition intensities to the macro-state *NRF*, corresponding to a non-repairable failure. This is only possible when a non-repairable failure occurs, and the repairperson is not at their workplace,

$$\mathbf{Q}^{NRF} = \begin{matrix} O^v \\ O^{nv} \\ RF \\ NRF \\ CR \\ PM \end{matrix} \begin{pmatrix} 0 & 0 & 0 & \mathbf{Q}^{NRF}_{O^v,NRF} & 0 & 0 \\ 0 & 0 & 0 & 0 & 0 & 0 \\ 0 & 0 & 0 & 0 & 0 & 0 \\ 0 & 0 & 0 & 0 & 0 & 0 \\ 0 & 0 & 0 & 0 & 0 & 0 \\ 0 & 0 & 0 & 0 & 0 & 0 \end{pmatrix},$$

where

$\mathbf{Q}^{NRF}_{O^v,RF} = \mathbf{H}_{NRF}(\mathbf{U}_{1,\ldots,K} = \mathbf{I}, 1, 1) \otimes \mathbf{I}$.

A non-repairable failure occurs regardless of the internal level of the system, with no new unit and no additional wear due to external shocks ($\mathbf{H}_{RF}(\mathbf{I}, 1, 1)$). There is no modification in transitions during vacation time ($\mathbf{I}$).

Matrix $\mathbf{Q}^{R}$

This matrix contains the transition intensities to the macro-state $O^{nv}$, transitioning from operability with the repairperson on vacation to operability with the repairperson at their workplace. This is only possible when the repairperson returns and, probabilistically, depending on the operational level, decides to remain at their workplace,

$$\mathbf{Q}^{R} = \begin{matrix} O^v \\ O^{nv} \\ RF \\ NRF \\ CR \\ PM \end{matrix} \begin{pmatrix} 0 & \mathbf{Q}^{R}_{O^v,O^{nv}} & 0 & 0 & 0 & 0 \\ 0 & 0 & 0 & 0 & 0 & 0 \\ 0 & 0 & 0 & 0 & 0 & 0 \\ 0 & 0 & 0 & 0 & 0 & 0 \\ 0 & 0 & 0 & 0 & 0 & 0 \\ 0 & 0 & 0 & 0 & 0 & 0 \end{pmatrix},$$

where

$$\mathbf{Q}^{R}_{O^v,O^{nv}} = \sum_{k=1}^{K-1} \mathbf{U}_k \mathbf{U}'_{1,\ldots,K-1} \otimes \mathbf{I} \otimes \mathbf{I} \otimes \mathbf{V}^0 (1 - p_k).$$

The system is in a state prior to *K*; otherwise, upon the repairperson's return, it would transition to preventive maintenance ($\sum_{k=1}^{K-1} \mathbf{U}_k$). After the repairperson is incorporated, the system remains at its current level, $\sum_{k=1}^{K-1} \mathbf{U}_k \mathbf{U}'_{1,\ldots,K-1}$. Both the external shock time and the wear remain unchanged, $\sum_{k=1}^{K-1} \mathbf{U}_k \mathbf{U}'_{1,\ldots,K-1} \otimes \mathbf{I} \otimes \mathbf{I}$. Finally, the repairperson has been incorporated and remains at their workstation with the probability indicated according to the degradation level, $\mathbf{V}^0 (1 - p_k)$.

Matrix $\mathbf{Q}^{R+PM}$

This matrix contains the transition intensities to the macro-state *PM* when the repairperson returns from the vacation period. For this to occur, the repairperson returns and observes a critical operating level,

$$\mathbf{Q}^{R+PM} = \begin{matrix} O^v \\ O^{nv} \\ RF \\ NRF \\ CR \\ PM \end{matrix} \begin{pmatrix} 0 & 0 & 0 & 0 & 0 & \mathbf{Q}^{R+PM}_{O^v,PM} \\ 0 & 0 & 0 & 0 & 0 & 0 \\ 0 & 0 & 0 & 0 & 0 & 0 \\ 0 & 0 & 0 & 0 & 0 & 0 \\ 0 & 0 & 0 & 0 & 0 & 0 \\ 0 & 0 & 0 & 0 & 0 & 0 \end{pmatrix},$$

where

$\mathbf{Q}^{R+PM}_{O^v,PM} = \mathbf{U}_K \mathbf{e} \otimes \mathbf{I} \otimes \mathbf{e} \otimes \mathbf{V}^0 \otimes \boldsymbol{\beta}^2$.

The system is operational, but the repairperson is not at their workplace. The only transition that occurs is the arrival of the repairer, $\mathbf{V}^0$, who observes a critical operating level, $\mathbf{U}_K$, and begins preventive maintenance, $\boldsymbol{\beta}^2$. There is no transition in the external failure time, $\mathbf{I}$.

Matrix $\mathbf{Q}^{R+NVP}$

This matrix is considered when the system is operational without the repairperson in the repair channel, this one returns, and then departs again for another vacation period,

$$\mathbf{Q}^{R+NVP} = \begin{matrix} O^v \\ O^{nv} \\ RF \\ NRF \\ CR \\ PM \end{matrix} \begin{pmatrix} \mathbf{Q}^{R+NVP}_{O^v,O^v} & \mathbf{0} & \mathbf{0} & \mathbf{0} & \mathbf{0} & \mathbf{0} \\ \mathbf{0} & \mathbf{0} & \mathbf{0} & \mathbf{0} & \mathbf{0} & \mathbf{0} \\ \mathbf{0} & \mathbf{0} & \mathbf{0} & \mathbf{0} & \mathbf{0} & \mathbf{0} \\ \mathbf{0} & \mathbf{0} & \mathbf{0} & \mathbf{0} & \mathbf{0} & \mathbf{0} \\ \mathbf{0} & \mathbf{0} & \mathbf{0} & \mathbf{0} & \mathbf{0} & \mathbf{0} \\ \mathbf{0} & \mathbf{0} & \mathbf{0} & \mathbf{0} & \mathbf{0} & \mathbf{0} \end{pmatrix},$$

where

$$\mathbf{Q}^{R+NVP}_{O^v,O^v} = \sum_{k=1}^{K-1} \mathbf{U}_k \otimes \mathbf{I} \otimes \mathbf{I} \otimes p_k \mathbf{V}^0 \boldsymbol{v}.$$

Once again, the only transition that occurs is the repairperson's return and the beginning of a new vacation period, $\mathbf{V}^0 \boldsymbol{v}$. For this to happen, the repairperson must observe a non-critical operating level, $\mathbf{U}_k$ and leave with probability $p_k$. This applies to any non-critical level, where $k = 1,\ldots, K-1$.

Matrix $\mathbf{Q}^{R+CR}$

The transitions considered in this case occur when the system is non-operational due to a reparable failure, the repairperson returns, and corrective repair begins,

$$\mathbf{Q}^{R+CR} = \begin{matrix} O^v \\ O^{nv} \\ RF \\ NRF \\ CR \\ PM \end{matrix} \begin{pmatrix} \mathbf{0} & \mathbf{0} & \mathbf{0} & \mathbf{0} & \mathbf{0} & \mathbf{0} \\ \mathbf{0} & \mathbf{0} & \mathbf{0} & \mathbf{0} & \mathbf{0} & \mathbf{0} \\ \mathbf{0} & \mathbf{0} & \mathbf{0} & \mathbf{0} & \mathbf{Q}^{R+CR}_{RF,CR} & \mathbf{0} \\ \mathbf{0} & \mathbf{0} & \mathbf{0} & \mathbf{0} & \mathbf{0} & \mathbf{0} \\ \mathbf{0} & \mathbf{0} & \mathbf{0} & \mathbf{0} & \mathbf{0} & \mathbf{0} \\ \mathbf{0} & \mathbf{0} & \mathbf{0} & \mathbf{0} & \mathbf{0} & \mathbf{0} \end{pmatrix},$$

where

$$\mathbf{Q}^{R+CR}_{RF,CR} = \mathbf{I} \otimes \mathbf{V}^0 \otimes \boldsymbol{\beta}^1.$$

The only transition is the repairperson's return and the probabilistic initiation of corrective repair, $\mathbf{V}^0 \otimes \boldsymbol{\beta}^1$. The external failure time remains unchanged, $\mathbf{I}$.

Matrix $\mathbf{Q}^{R+NU}$

The system is in a non-repairable failure state, and upon the repairperson's return, the unit is renewed, and a new vacation period begins,

$$\mathbf{Q}^{R+NU} = \begin{matrix} O^v \\ O^{nv} \\ RF \\ NRF \\ CR \\ PM \end{matrix} \begin{pmatrix} \mathbf{0} & \mathbf{0} & \mathbf{0} & \mathbf{0} & \mathbf{0} & \mathbf{0} \\ \mathbf{0} & \mathbf{0} & \mathbf{0} & \mathbf{0} & \mathbf{0} & \mathbf{0} \\ \mathbf{0} & \mathbf{0} & \mathbf{0} & \mathbf{0} & \mathbf{0} & \mathbf{0} \\ \mathbf{Q}^{R+NU}_{NRF,O^v} & \mathbf{0} & \mathbf{0} & \mathbf{0} & \mathbf{0} & \mathbf{0} \\ \mathbf{0} & \mathbf{0} & \mathbf{0} & \mathbf{0} & \mathbf{0} & \mathbf{0} \\ \mathbf{0} & \mathbf{0} & \mathbf{0} & \mathbf{0} & \mathbf{0} & \mathbf{0} \end{pmatrix},$$

where

$$\mathbf{Q}^{R+NU}_{NRF,O^v} = [\boldsymbol{\alpha} \otimes \mathbf{I} \otimes \boldsymbol{\omega}] \otimes \mathbf{V}^0 \boldsymbol{v}.$$

The repairperson's return occurs with the initiation of a new vacation period, $\mathbf{V}^0 \boldsymbol{v}$, as the repairperson observes that the unit is broken. A new unit is incorporated with an initial distribution, $\boldsymbol{\alpha} \otimes \mathbf{I} \otimes \boldsymbol{\omega}$.

Matrix $\mathbf{Q}^O$

The matrix $\mathbf{Q}^\circ$ contains the transitions that do not account for the events of the MMAP described in Section 3.3,

$$\mathbf{Q}^O = \begin{matrix} O^v \\ O^{nv} \\ RF \\ NRF \\ CR \\ PM \end{matrix} \begin{pmatrix} \mathbf{Q}^O_{O^v,O^v} & \mathbf{0} & \mathbf{0} & \mathbf{0} & \mathbf{0} & \mathbf{0} \\ \mathbf{0} & \mathbf{Q}^O_{O^{nv},O^{nv}} & \mathbf{0} & \mathbf{0} & \mathbf{0} & \mathbf{0} \\ \mathbf{0} & \mathbf{0} & \mathbf{Q}^O_{RF,RF} & \mathbf{0} & \mathbf{0} & \mathbf{0} \\ \mathbf{0} & \mathbf{0} & \mathbf{0} & \mathbf{Q}^O_{NRF,NRF} & \mathbf{0} & \mathbf{0} \\ \mathbf{Q}^O_{CR,O^v} & \mathbf{0} & \mathbf{0} & \mathbf{0} & \mathbf{Q}^O_{CR,CR} & \mathbf{0} \\ \mathbf{Q}^O_{PM,O^v} & \mathbf{0} & \mathbf{0} & \mathbf{0} & \mathbf{0} & \mathbf{Q}^O_{PM,PM} \end{pmatrix},$$

where

$$\mathbf{Q}^O_{O^v,O^v} = \mathbf{H}_O(\mathbf{I},\mathbf{I},\mathbf{I}) \otimes \mathbf{I} + \mathbf{I} \otimes \mathbf{I} \otimes \mathbf{I} \otimes \mathbf{V},$$

$$\mathbf{Q}^O_{O^{nv},O^{nv}} = \mathbf{H}_O\left(\mathbf{U}_{1,\ldots,K-1},\mathbf{U}'_{1,\ldots,K-1},\mathbf{I}\right),$$

$$\mathbf{Q}^O_{RF,RF} = \mathbf{I} \otimes \mathbf{V} + \left(\mathbf{L} + \mathbf{L}^0 \boldsymbol{\gamma}\right) \otimes \mathbf{I},$$

$$\mathbf{Q}^O_{NRF,NRF} = \mathbf{I} \otimes \mathbf{V} + \left(\mathbf{L} + \mathbf{L}^0 \boldsymbol{\gamma}\right) \otimes \mathbf{I},$$

$$\mathbf{Q}^O_{CR,CR} = \mathbf{I} \otimes \mathbf{S}_1 + \left(\mathbf{L} + \mathbf{L}^0 \boldsymbol{\gamma}\right) \otimes \mathbf{I},$$

$$\mathbf{Q}^O_{PM,PM} = \mathbf{I} \otimes \mathbf{S}_2 + \left(\mathbf{L} + \mathbf{L}^0 \boldsymbol{\gamma}\right) \otimes \mathbf{I},$$

$$\mathbf{Q}^O_{CR,O^v} = \boldsymbol{\alpha} \otimes \mathbf{I} \otimes \boldsymbol{\omega} \otimes \mathbf{S}^0_1 \otimes \boldsymbol{v},$$

$$\mathbf{Q}^O_{PM,O^v} = \boldsymbol{\alpha} \otimes \mathbf{I} \otimes \boldsymbol{\omega} \otimes \mathbf{S}^0_2 \otimes \boldsymbol{v}.$$

## APPENDIX B1

*Repairable Failure Function for the discrete case ($H_{RF}$)*

We assume the unit undergoes a repairable failure at a certain time this event can occur by (similar reasoning to the continuous case can be done for the matrix **U**):

- A repairable failure occurs because there is an internal failure and there is no external shock, $\mathbf{U}\mathbf{T}^0_r \otimes \mathbf{L} \otimes \mathbf{e}$.
- An external shock does not cause an irreparable failure but alters the internal performance, leading to an internal failure. Alternatively, the system may experience an internal failure independent of the shock, $\mathbf{U}(\mathbf{T}^0_r + \mathbf{T}\mathbf{W}^0_r) \otimes \mathbf{L}^0 \boldsymbol{\gamma}(1-\omega^0) \otimes \mathbf{C}\mathbf{e}$.

Then the matrix function for this event is

$$\mathbf{H}_{RF}(\mathbf{U}) = \mathbf{U}\mathbf{T}^0_r \otimes \mathbf{L} \otimes \mathbf{e} + \mathbf{U}(\mathbf{T}^0_r + \mathbf{T}\mathbf{W}^0_r) \otimes \mathbf{L}^0 \boldsymbol{\gamma}(1-\omega^0) \otimes \mathbf{C}\mathbf{e}.$$

*Non-Repairable Failure Function for the Discrete Case ($H_{NRF}$)*

The unit may undergo a non-repairable failure because (similar reasoning to the continuous case can be done for the matrix **U**, **R** and **A**):

- An internal non-repairable failure occurs with no external shock, $\mathbf{U}\mathbf{T}^0_{nr}\mathbf{R} \otimes \mathbf{L} \otimes \mathbf{e}\mathbf{A}$.
- An external shock occurs but does not provoke total failure $(1-\omega^0)$. This shock provokes a non-repairable internal failure or, irrespective of the shock, the unit may experience a non-repairable failure, $\mathbf{U}(\mathbf{T}^0_{nr} + \mathbf{T}\mathbf{W}^0_{nr})\mathbf{R} \otimes \mathbf{L}^0 \boldsymbol{\gamma}(1-\omega^0) \otimes \mathbf{C}\mathbf{e}\mathbf{A}$.
- An external shock causes a non-repairable total failure, $\mathbf{U}\mathbf{e}\mathbf{R} \otimes \mathbf{L}^0 \boldsymbol{\gamma}\omega^0$.
- There is an external shock but does not provoke the non-repairable failure; this external shock causes a modification in the external cumulative damage causing the non-repairable failure, $\mathbf{U}\mathbf{e}\mathbf{S} \otimes \mathbf{L}^0 \boldsymbol{\gamma}(1-\omega^0) \otimes \mathbf{C}^0 \mathbf{Q}$.

Therefore, the matrix that govern this transition is given by:

$$\mathbf{H}_{NRF}(\mathbf{U},\mathbf{R},\mathbf{A}) = \mathbf{U}\mathbf{T}^0_{nr}\mathbf{R} \otimes \mathbf{L} \otimes \mathbf{e}\mathbf{A} + \mathbf{U}(\mathbf{T}^0_{nr} + \mathbf{T}\mathbf{W}^0_{nr})\mathbf{R} \otimes \mathbf{L}^0 \boldsymbol{\gamma}(1-\omega^0) \otimes \mathbf{C}\mathbf{e}\mathbf{A}$$
$$+\mathbf{U}\mathbf{e}\mathbf{R} \otimes \mathbf{L}^0 \boldsymbol{\gamma}\omega^0 \otimes \mathbf{e}\mathbf{A} + \mathbf{U}\mathbf{e}\mathbf{R} \otimes \mathbf{L}^0 \boldsymbol{\gamma}(1-\omega^0) \otimes \mathbf{C}^0 \mathbf{A}.$$

*No Events at a Certain Time Function for the Discrete Case ($H_O$)*

We assume that the unit is operational and this time continues working. This occurs because by different situations (similar reasoning to the continuous case can be done for the matrix **U**, **R** and **A**):

- The internal performance continues in the same level or changes to another, equally operational state. The repairperson continues on vacation. There is no external shock occurs, $\mathbf{U}\mathbf{T}\mathbf{R} \otimes \mathbf{L} \otimes \mathbf{A}$.
- There is an external shock, but does not occur a total failure, this unit might modify the internal performance but does not produce internal failure, and the repairperson continues on vacation, $\mathbf{U}\mathbf{T}\mathbf{W}\mathbf{R} \otimes \mathbf{L}^0 \boldsymbol{\gamma}(1-\omega^0) \otimes \mathbf{C}\mathbf{A}$.

Then,
$$\mathbf{H}_O(\mathbf{U},\mathbf{R},\mathbf{A}) = \mathbf{U}\mathbf{T}\mathbf{R} \otimes \mathbf{L} \otimes \mathbf{A} + \mathbf{U}\mathbf{T}\mathbf{W}\mathbf{R} \otimes \mathbf{L}^0 \boldsymbol{\gamma}(1-\omega^0) \otimes \mathbf{C}\mathbf{A}.$$

## APPENDIX B2

This appendix contains the remaining block matrices for the system in the discrete case described in Section 3.4.1. Unlike the continuous case, in the discrete case, multiple events, whether conditioned or not, can occur simultaneously.

Matrices $\mathbf{D}^{RF+CR}$ and $\mathbf{D}^{NRF+NU}$

The general structure of the matrices $\mathbf{D}^{RF+CR}$ and $\mathbf{D}^{NRF+NU}$ is similar to that of the continuous case. Thus,

$$\mathbf{D}^{RF+CR} = \begin{array}{c} O^v \\ O^{nv} \\ RF \\ NRF \\ CR \\ PM \end{array} \begin{pmatrix} 0 & 0 & 0 & 0 & 0 & 0 \\ 0 & 0 & 0 & 0 & \mathbf{D}^{RF+CR}_{O^{nv},CR} & 0 \\ 0 & 0 & 0 & 0 & 0 & 0 \\ 0 & 0 & 0 & 0 & 0 & 0 \\ 0 & 0 & 0 & 0 & 0 & 0 \\ 0 & 0 & 0 & 0 & 0 & 0 \end{pmatrix}, \quad \mathbf{D}^{NRF+NU} = \begin{array}{c} O^v \\ O^{nv} \\ RF \\ NRF \\ CR \\ PM \end{array} \begin{pmatrix} 0 & 0 & 0 & 0 & 0 & 0 \\ \mathbf{D}^{NRF+NU}_{O^{nv},O^v} & 0 & 0 & 0 & 0 & 0 \\ 0 & 0 & 0 & 0 & 0 & 0 \\ \vdots & \vdots & \vdots & \vdots & \vdots & \vdots \\ \vdots & \vdots & \vdots & \vdots & \vdots & \vdots \\ 0 & 0 & 0 & 0 & 0 & 0 \end{pmatrix}$$

where

$$\mathbf{D}^{RF+CR}_{O^{nv},CR} = \mathbf{H}_{RF}\left(\mathbf{U}_{1,\ldots,K-1}\right) \otimes \boldsymbol{\beta}^1, \mathbf{D}^{NRF+NU}_{O^{nv},O^v} = \mathbf{H}_{NRF}\left(\mathbf{U}_{1,\ldots,K-1},\boldsymbol{\alpha},\boldsymbol{\omega}\right) \otimes \boldsymbol{\upsilon}$$

Matrix $\mathbf{D}^{RF}$

The $\mathbf{D}^{RF}$ matrix contains the transition probabilities when the system is operational, and a repairable failure occurs. This situation is only possible when the repairer is on vacation, and a repairable failure takes place,

$$\mathbf{D}^{RF} = \begin{array}{c} O^v \\ O^{nv} \\ RF \\ NRF \\ CR \\ PM \end{array} \begin{pmatrix} 0 & 0 & \mathbf{D}^{RF}_{O^v,RF} & 0 & 0 & 0 \\ 0 & 0 & 0 & 0 & 0 & 0 \\ 0 & 0 & 0 & 0 & 0 & 0 \\ 0 & 0 & 0 & 0 & 0 & 0 \\ 0 & 0 & 0 & 0 & 0 & 0 \\ 0 & 0 & 0 & 0 & 0 & 0 \end{pmatrix},$$

where

$$\mathbf{D}^{RF}_{O^v,RF} = \mathbf{H}_{RF}\left(\mathbf{U}_{1,\ldots,K}\right) \otimes \mathbf{V}.$$

The system is operational, and a repairable failure occurs in the unit, whether due to a shock or not ($\mathbf{H}_{RF}\left(\mathbf{U}_{1,\ldots,K} = \mathbf{I}\right)$). Simultaneously, a transition in the repairer's vacation time without returning ($\mathbf{V}$) may also occur.

Matrix $\mathbf{D}^{NRF}$

The $\mathbf{D}^{NRF}$ matrix contains the transition probabilities when the system is operational, and a non-repairable failure occurs. This situation is only possible when the repairer is on vacation, and a non-repairable failure takes place,

$$\mathbf{D}^{NRF} = \begin{array}{c} O^v \\ O^{nv} \\ RF \\ NRF \\ CR \\ PM \end{array} \begin{pmatrix} 0 & 0 & 0 & \mathbf{D}^{NRF}_{O^v,NRF} & 0 & 0 \\ 0 & 0 & 0 & 0 & 0 & 0 \\ 0 & 0 & 0 & 0 & 0 & 0 \\ 0 & 0 & 0 & 0 & 0 & 0 \\ 0 & 0 & 0 & 0 & 0 & 0 \\ 0 & 0 & 0 & 0 & 0 & 0 \end{pmatrix},$$

where

$$\mathbf{D}^{NRF}_{O^v,NRF} = \mathbf{H}_{NRF}\left(\mathbf{U}_{1,\ldots,K} = \mathbf{I}, 1, 1\right) \otimes \mathbf{V}.$$

The system is operational, and a non-repairable failure occurs in the unit, either due to external causes such as shocks or internal issues ($\mathbf{H}_{NRF}\left(\mathbf{U}_{1,\ldots,K} = \mathbf{I}, 1, 1\right)$). Simultaneously, a transition in the repairer's vacation time without returning ($\mathbf{V}$) may also occur.

Matrix $\mathbf{D}^R$

The $\mathbf{D}^R$ matrix contains the transition probabilities when the system is operational, and the repairer returns, remaining in the repair channel waiting for a failure while the system continues to operate,

$$\mathbf{D}^R = \begin{array}{c} O^v \\ O^{nv} \\ RF \\ NRF \\ CR \\ PM \end{array} \begin{pmatrix} 0 & \mathbf{D}^R_{O^v,O^{nv}} & 0 & 0 & 0 & 0 \\ 0 & 0 & 0 & 0 & 0 & 0 \\ 0 & 0 & 0 & 0 & 0 & 0 \\ 0 & 0 & 0 & 0 & 0 & 0 \\ 0 & 0 & 0 & 0 & 0 & 0 \\ 0 & 0 & 0 & 0 & 0 & 0 \end{pmatrix},$$

where

$$\mathbf{D}^R_{O^v,O^{nv}} = \sum_{k=1}^{K-1} \mathbf{H}_O\left(\mathbf{U}_k, \mathbf{U}'_{1,\ldots,K-1}, \mathbf{I}\right) \otimes (1-p_k)\mathbf{V}^0.$$

In this case, from any level $k$, a failure-free transition occurs to any level below the critical level $K$, and with or without external shock the internal is in a level below than $K$ ($\mathbf{H}_O\left(\mathbf{U}_k, \mathbf{U}'_{1,\ldots,K-1}, \mathbf{I}\right)$). The repairperson returns ($\mathbf{V}^0$) and observes no critical damage, and no failures. He stays in the repair channel with a probability of $1 - p_k$.

Matrix $\mathbf{D}^{R+PM}$

The $\mathbf{D}^{R+PM}$ matrix contains the transition probabilities when the system is operational, and the repairperson returns observing critical level and initiates preventive maintenance,

$$\mathbf{D}^{R+PM} = \begin{array}{c} O^v \\ O^{nv} \\ RF \\ NRF \\ CR \\ PM \end{array} \begin{pmatrix} 0 & 0 & 0 & 0 & 0 & \mathbf{D}^{R+PM}_{O^v,PM} \\ 0 & 0 & 0 & 0 & 0 & 0 \\ 0 & 0 & 0 & 0 & 0 & 0 \\ 0 & 0 & 0 & 0 & 0 & 0 \\ 0 & 0 & 0 & 0 & 0 & 0 \\ 0 & 0 & 0 & 0 & 0 & 0 \end{pmatrix},$$

where

$\mathbf{D}^{R+PM}_{O^v,PM} = \mathbf{H}_O(\mathbf{I}, \mathbf{U}_K \mathbf{e}, \mathbf{e}) \otimes \mathbf{V}^0 \otimes \boldsymbol{\beta}^2.$

The system remains operational without any failure in the unit, ($\mathbf{H}_O(\mathbf{I}, \mathbf{U}_K \mathbf{e}, \mathbf{e})$), the repairperson returns, observes critical level and preventive maintenance begins ($\mathbf{V}^0 \otimes \boldsymbol{\beta}^2$).

Matrix $\mathbf{D}^{R+NVP}$

The $\mathbf{D}^{R+NVP}$ matrix contains the transition probabilities when the system is operational with the repairperson on vacation. Upon returning, the repairperson observes the system's state (neither a failure nor a critical level) and begins another vacation period,

$$\mathbf{D}^{R+NVP} = \begin{array}{c} O^v \\ O^{nv} \\ RF \\ NRF \\ CR \\ PM \end{array} \begin{pmatrix} \mathbf{D}^{R+NVP}_{O^v,O^v} & 0 & 0 & 0 & 0 & 0 \\ 0 & 0 & 0 & 0 & 0 & 0 \\ 0 & 0 & 0 & 0 & 0 & 0 \\ 0 & 0 & 0 & 0 & 0 & 0 \\ 0 & 0 & 0 & 0 & 0 & 0 \\ 0 & 0 & 0 & 0 & 0 & 0 \end{pmatrix},$$

where

$\mathbf{D}^{R+NVP}_{O^v,O^v} = \sum_{k=1}^{K-1} \mathbf{H}_O(\mathbf{I}, \mathbf{U}_k, \mathbf{I}) \otimes p_k \mathbf{V}^0 \boldsymbol{v}.$

From any level, a failure-free transition occurs to any level below the critical level $K$ ($\mathbf{H}_O(\mathbf{I}, \mathbf{U}_k, \mathbf{I})$). The repairperson returns, observes the system, and with probability $p_k$, begins another vacation period ($p_k \mathbf{V}^0 \boldsymbol{v}$).

Matrix $\mathbf{D}^{R+CR}$

The $\mathbf{D}^{R+CR}$ matrix contains the transition probabilities when the repairperson returns and observes that previously the system underwent a repairable failure, initiating corrective maintenance,

$$\mathbf{D}^{R+CR} = \begin{array}{c} O^v \\ O^{nv} \\ RF \\ NRF \\ CR \\ PM \end{array} \begin{pmatrix} 0 & 0 & 0 & 0 & 0 & 0 \\ 0 & 0 & 0 & 0 & 0 & 0 \\ 0 & 0 & 0 & 0 & \mathbf{D}^{R+CR}_{RF,CR} & 0 \\ 0 & 0 & 0 & 0 & 0 & 0 \\ 0 & 0 & 0 & 0 & 0 & 0 \\ 0 & 0 & 0 & 0 & 0 & 0 \end{pmatrix},$$

where

$\mathbf{D}^{R+CR}_{RF,CR} = \left(\mathbf{L} + \mathbf{L}^0 \boldsymbol{\gamma}\right) \otimes \mathbf{V}^0 \otimes \boldsymbol{\beta}^1.$

Regardless of the external shock factor (which persists even in the absence of an operational unit), the repairperson returns and begins corrective maintenance.

Matrix $\mathbf{D}^{R+NU}$

The $\mathbf{D}^{R+NU}$ matrix contains the transition probabilities when the repairperson returns, and observes previous non-repairable failure. The unit is replaced. Then,

$$\mathbf{D}^{R+NU} = \begin{array}{c} O^v \\ O^{nv} \\ RF \\ NRF \\ CR \\ PM \end{array} \begin{pmatrix} \mathbf{0} & 0 & 0 & 0 & 0 & 0 \\ \mathbf{0} & 0 & 0 & 0 & 0 & 0 \\ \mathbf{0} & 0 & 0 & 0 & 0 & 0 \\ \mathbf{D}^{R+NU}_{NRF,O^v} & 0 & 0 & 0 & 0 & 0 \\ \mathbf{0} & 0 & 0 & 0 & 0 & 0 \\ \mathbf{0} & 0 & 0 & 0 & 0 & 0 \end{pmatrix}, \text{ where}$$

$$\mathbf{D}^{R+NU}_{NRF,O^v} = \left[\boldsymbol{\alpha} \otimes \left(\mathbf{L} + \mathbf{L}^0 \boldsymbol{\gamma}\right) \otimes \boldsymbol{\omega}\right] \otimes \mathbf{V}^0 \boldsymbol{v}.$$

Regardless of external behaviour ($\mathbf{L} + \mathbf{L}^0 \boldsymbol{\gamma}$), the repairer returns ($\mathbf{V}^0$), initiates a new unit ($\boldsymbol{\alpha}$), without initial external shock damage ($\boldsymbol{\omega}$), and begins a new vacation period ($\boldsymbol{v}$).

Matrix $\mathbf{D}^{R+RF+CR}$

The $\mathbf{D}^{R+RF+CF}$ matrix contains the transition probabilities when the system is operational, and simultaneously, a repairable failure occurs, and the repairer returns to begin corrective maintenance,

$$\mathbf{D}^{R+RF+CR} = \begin{array}{c} O^v \\ O^{nv} \\ RF \\ NRF \\ CR \\ PM \end{array} \begin{pmatrix} \mathbf{0} & \mathbf{0} & \mathbf{0} & \mathbf{0} & \mathbf{D}^{R+RF+CR}_{O^v,CR} & \mathbf{0} \\ \mathbf{0} & \mathbf{0} & \mathbf{0} & \mathbf{0} & \mathbf{0} & \mathbf{0} \\ \mathbf{0} & \mathbf{0} & \mathbf{0} & \mathbf{0} & \mathbf{0} & \mathbf{0} \\ \mathbf{0} & \mathbf{0} & \mathbf{0} & \mathbf{0} & \mathbf{0} & \mathbf{0} \\ \mathbf{0} & \mathbf{0} & \mathbf{0} & \mathbf{0} & \mathbf{0} & \mathbf{0} \\ \mathbf{0} & \mathbf{0} & \mathbf{0} & \mathbf{0} & \mathbf{0} & \mathbf{0} \end{pmatrix}$$

where

$$\mathbf{D}^{R+RF+CR}_{O^v,CR} = \mathbf{H}_{RF}\left(\mathbf{U}_{1,\ldots,K} = \mathbf{I}\right) \otimes \mathbf{V}^0 \otimes \boldsymbol{\beta}^1.$$

Matrix $\mathbf{D}^{R+NRF+NU}$

The $\mathbf{D}^{R+NRF+NU}$ matrix contains the transition probabilities when the system is operational, and simultaneously, a non-repairable failure occurs, and the repairperson returns to replace the unit,

$$\mathbf{D}^{R+NRF+NU} = \begin{array}{c} O^v \\ O^{nv} \\ RF \\ NRF \\ CR \\ PM \end{array} \begin{pmatrix} \mathbf{D}^{R+NRF+NU}_{O^v,O^v} & \mathbf{0} & \mathbf{0} & \mathbf{0} & \mathbf{0} & \mathbf{0} \\ \mathbf{0} & \mathbf{0} & \mathbf{0} & \mathbf{0} & \mathbf{0} & \mathbf{0} \\ \mathbf{0} & \mathbf{0} & \mathbf{0} & \mathbf{0} & \mathbf{0} & \mathbf{0} \\ \mathbf{0} & \mathbf{0} & \mathbf{0} & \mathbf{0} & \mathbf{0} & \mathbf{0} \\ \mathbf{0} & \mathbf{0} & \mathbf{0} & \mathbf{0} & \mathbf{0} & \mathbf{0} \\ \mathbf{0} & \mathbf{0} & \mathbf{0} & \mathbf{0} & \mathbf{0} & \mathbf{0} \end{pmatrix},$$

where

$$\mathbf{D}^{R+NRF+NU}_{O^v,O^v} = \mathbf{H}_{NRF}(\mathbf{I}, \boldsymbol{\alpha}, \boldsymbol{\omega}) \otimes \mathbf{V}^0 \boldsymbol{v}.$$

At the same time the repairperson returns ($\mathbf{V}^0$), a non-repairable failure occurs, and the unit is restarted, $\mathbf{H}_{NRF}(\mathbf{I}, \boldsymbol{\alpha}, \boldsymbol{\omega})$. The repair person begins a new vacation period, $\boldsymbol{v}$.

Matrix $\mathbf{D}^O$

The $\mathbf{D}^O$ matrix contains the transition probabilities that do not account for the MMAP events described in Section 3.4. The general structure of $\mathbf{D}^O$ is similar to that of the continuous case, $\mathbf{Q}^O$,

$$\mathbf{D}^O = \begin{array}{c} O^v \\ O^{nv} \\ RF \\ NRF \\ CR \\ PM \end{array} \begin{pmatrix} \mathbf{D}^O_{O^v,O^v} & \mathbf{0} & \mathbf{0} & \mathbf{0} & \mathbf{0} & \mathbf{0} \\ \mathbf{0} & \mathbf{D}^O_{O^{nv},O^{nv}} & \mathbf{0} & \mathbf{0} & \mathbf{0} & \mathbf{0} \\ \mathbf{0} & \mathbf{0} & \mathbf{D}^O_{RF,RF} & \mathbf{0} & \mathbf{0} & \mathbf{0} \\ \mathbf{0} & \mathbf{0} & \mathbf{0} & \mathbf{D}^O_{NRF,NRF} & \mathbf{0} & \mathbf{0} \\ \mathbf{D}^O_{CR,O^v} & \mathbf{0} & \mathbf{0} & \mathbf{0} & \mathbf{D}^O_{CR,CR} & \mathbf{0} \\ \mathbf{D}^O_{PM,O^v} & \mathbf{0} & \mathbf{0} & \mathbf{0} & \mathbf{0} & \mathbf{D}^O_{PM,PM} \end{pmatrix},$$

where

$$\mathbf{D}^O_{O^v,O^v} = \mathbf{H}_O(\mathbf{I},\mathbf{I},\mathbf{I}) \otimes \mathbf{V},$$

$$\mathbf{D}^O_{O^{nv},O^{nv}} = \mathbf{H}_O\left(\mathbf{U}_{1,\ldots,K-1}, \mathbf{U}'_{1,\ldots,K-1}, \mathbf{I}\right),$$

$$\mathbf{D}^O_{RF,RF} = \left(\mathbf{L} + \mathbf{L}^0\boldsymbol{\gamma}\right) \otimes \mathbf{V},$$

$$\mathbf{D}^O_{NRF,NRF} = \left(\mathbf{L} + \mathbf{L}^0\boldsymbol{\gamma}\right) \otimes \mathbf{V},$$

$$\mathbf{D}^O_{CR,CR} = \left(\mathbf{L} + \mathbf{L}^0\boldsymbol{\gamma}\right) \otimes \mathbf{S}_1,$$

$$\mathbf{D}^O_{PM,PM} = \left(\mathbf{L} + \mathbf{L}^0\boldsymbol{\gamma}\right) \otimes \mathbf{S}_2,$$

$$\mathbf{D}^O_{CR,O^v} = \boldsymbol{\alpha} \otimes \left(\mathbf{L} + \mathbf{L}^0\boldsymbol{\gamma}\right) \otimes \boldsymbol{\omega} \otimes \mathbf{S}_1 \otimes \boldsymbol{v},$$

$$\mathbf{D}^O_{PM,O^v} = \boldsymbol{\alpha} \otimes \left(\mathbf{L} + \mathbf{L}^0\boldsymbol{\gamma}\right) \otimes \boldsymbol{\omega} \otimes \mathbf{S}_2 \otimes \boldsymbol{v}.$$

**APPENDIX B3**

The stationary distribution for the discrete case is developed as follows from the balance equations and the normalization condition, $\pi D = \pi$, $\pi e = 1$.

The transition probability matrix is expressed as,

$$\mathbf{D} = \begin{array}{c} O^v \\ O^{nv} \\ RF \\ NRF \\ CR \\ PM \end{array} \begin{pmatrix} \mathbf{D}_{11} & \mathbf{D}_{12} & \mathbf{D}_{13} & \mathbf{D}_{14} & \mathbf{D}_{15} & \mathbf{D}_{16} \\ \mathbf{D}_{21} & \mathbf{D}_{22} & \mathbf{0} & \mathbf{0} & \mathbf{D}_{25} & \mathbf{D}_{26} \\ \mathbf{0} & \mathbf{0} & \mathbf{D}_{33} & \mathbf{0} & \mathbf{D}_{35} & \mathbf{0} \\ \mathbf{D}_{41} & \mathbf{0} & \mathbf{0} & \mathbf{D}_{44} & \mathbf{0} & \mathbf{0} \\ \mathbf{D}_{51} & \mathbf{0} & \mathbf{0} & \mathbf{0} & \mathbf{D}_{55} & \mathbf{0} \\ \mathbf{D}_{61} & \mathbf{0} & \mathbf{0} & \mathbf{0} & \mathbf{0} & \mathbf{D}_{66} \end{pmatrix},$$

where

$$\mathbf{D}_{11} = \mathbf{D}^{R+NRF+NU}_{O^v,O^v} + \mathbf{D}^{R+NVP}_{O^v,O^v} + \mathbf{D}^O_{O^v,O^v},$$

$$\mathbf{D}_{15} = \mathbf{D}^{R+RF+CR}_{O^v,CR}$$

$$\mathbf{D}_{26} = \mathbf{D}^{PM}_{O^{nv},PM}; \mathbf{D}_{33} = \mathbf{D}^{O}_{RF,RF}; \mathbf{D}_{35} = \mathbf{D}^{R+CR}_{RF,CR}; \mathbf{D}_{41} = \mathbf{D}^{R+NU}_{NRF,O^v}; \mathbf{D}_{44} = \mathbf{D}^{O}_{NRF,NRF}; \mathbf{D}_{51} = \mathbf{D}^{O}_{CR,O^v};$$

$$\mathbf{D}_{55} = \mathbf{D}^O_{CR,CR}; \mathbf{D}_{61} = \mathbf{D}^O_{PM,O^v}; \mathbf{D}_{66} = \mathbf{D}^O_{PM,PM}.$$

In a matrix way, the balance equations are given by

$$\pi_1 \mathbf{D}_{11} + \pi_2 \mathbf{D}_{21} + \pi_4 \mathbf{D}_{41} + \pi_5 \mathbf{D}_{51} + \pi_6 \mathbf{D}_{61} = \pi_1$$
$$\pi_1 \mathbf{D}_{12} + \pi_2 \mathbf{D}_{22} = \pi_2$$
$$\pi_1 \mathbf{D}_{13} + \pi_3 \mathbf{D}_{33} = \pi_3$$
$$\pi_1 \mathbf{D}_{14} + \pi_4 \mathbf{D}_{44} = \pi_4$$
$$\pi_1 \mathbf{D}_{15} + \pi_2 \mathbf{D}_{25} + \pi_3 \mathbf{D}_{35} + \pi_5 \mathbf{D}_{55} = \pi_5$$
$$\pi_1 \mathbf{D}_{16} + \pi_2 \mathbf{D}_{26} + \pi_6 \mathbf{D}_{66} = \pi_6$$
$$\pi e = 1.$$

Therefore

$$\pi_2 = \pi_1 \mathbf{H}_{12}; \pi_3 = \pi_1 \mathbf{H}_{13}; \pi_4 = \pi_1 \mathbf{H}_{14}; \pi_5 = \pi_1 \mathbf{H}_{15}; \pi_6 = \pi_1 \mathbf{H}_{16}, \tag{14}$$

where

$$\mathbf{H}_{12} = \mathbf{D}_{12}(\mathbf{I} - \mathbf{D}_{22})^{-1}; \mathbf{H}_{13} = \mathbf{D}_{13}(\mathbf{I} - \mathbf{D}_{33})^{-1}; \mathbf{H}_{14} = \mathbf{D}_{14}(\mathbf{I} - \mathbf{D}_{44})^{-1};$$

$$\mathbf{H}_{15} = (\mathbf{D}_{15} + \mathbf{H}_{12}\mathbf{D}_{25} + \mathbf{H}_{13}\mathbf{D}_{35})(\mathbf{I} - \mathbf{D}_{55})^{-1}; \mathbf{H}_{16} = (\mathbf{D}_{16} + \mathbf{H}_{12}\mathbf{D}_{26})(\mathbf{I} - \mathbf{D}_{66})^{-1}.$$

Thus, from the first equation and the normalization condition,

$$\pi_1 \mathbf{D}_{11} + \pi_1 \mathbf{H}_{12}\mathbf{D}_{21} + \pi_1 \mathbf{H}_{14}\mathbf{D}_{41} + \pi_1 \mathbf{H}_{15}\mathbf{D}_{51} + \pi_1 \mathbf{H}_{16}\mathbf{D}_{61} - \pi_1 = \mathbf{0}$$

$$\pi_1 \mathbf{e} + \pi_1 \mathbf{H}_{12}\mathbf{e} + \pi_1 \mathbf{H}_{13}\mathbf{e} + \pi_1 \mathbf{H}_{14}\mathbf{e} + \pi_1 \mathbf{H}_{15}\mathbf{e} + \pi_1 \mathbf{H}_{16}\mathbf{e} = 1$$

$$\pi_1 (\mathbf{e} + \mathbf{H}_{12}\mathbf{e} + \mathbf{H}_{13}\mathbf{e} + \mathbf{H}_{14}\mathbf{e} + \mathbf{H}_{15}\mathbf{e} + \mathbf{H}_{16}\mathbf{e}) = 1.$$

Therefore,

$$\pi_1 (\mathbf{D}_{11} + \mathbf{H}_{12}\mathbf{D}_{21} + \mathbf{H}_{14}\mathbf{D}_{41} + \mathbf{H}_{15}\mathbf{D}_{51} + \mathbf{H}_{16}\mathbf{D}_{61} - \mathbf{I}) = \mathbf{0},$$

$$\pi_1 (\mathbf{e} + \mathbf{H}_{12}\mathbf{e} + \mathbf{H}_{13}\mathbf{e} + \mathbf{H}_{14}\mathbf{e} + \mathbf{H}_{15}\mathbf{e} + \mathbf{H}_{16}\mathbf{e}) = 1.$$

Then,

$$\begin{aligned}\boldsymbol{\pi}_1 = (1,\mathbf{0})[\mathbf{e} + \mathbf{H}_{12}\mathbf{e} + \mathbf{H}_{13}\mathbf{e} + \mathbf{H}_{14}\mathbf{e} + \mathbf{H}_{15}\mathbf{e} + \mathbf{H}_{16}\mathbf{e}|\\(\mathbf{D}_{11} + \mathbf{H}_{12}\mathbf{D}_{21} + \mathbf{H}_{14}\mathbf{D}_{41} + \mathbf{H}_{15}\mathbf{D}_{51} + \mathbf{H}_{16}\mathbf{D}_{61} - \mathbf{I}) * ]^{-1}.\end{aligned} \quad (15)$$

### Data availability

No data was used for the research described in the article.